\input amstex
\documentstyle{amsppt}
\magnification=1200
\font\ftitle=cmbx12
\font\fsection=cmbx12
\line{}
\vskip4.5truecm
\NoBlackBoxes

\topmatter
\centerline{\ftitle An Analogue of the Kac-Wakimoto Formula}
\smallskip
\centerline{\ftitle and Black Hole Conditional Entropy}
\bigskip\smallskip
\centerline{{\smc  Roberto Longo} \footnote"$^*$"{Supported
in part by MURST and
CNR-GNAFA.\hfill\break
 E-mail  longo\@mat.utovrm.it
}} \bigskip \centerline{ Dipartimento di
Matematica, Universit\`a di Roma ``Tor Vergata''}
\centerline {via della Ricerca Scientifica, I--00133 Roma, Italy}
\medskip\centerline{Centro Linceo Interdisciplinare, Accademia
Nazionale dei Lincei} \centerline{via della Lungara 10, I--00165 Roma,
Italy} \vskip 2truecm

{\it Abstract.} A local formula for the dimension of
a superselection sector in Quantum Field Theory is obtained as
vacuum expectation value of the exponential of the proper
Hamiltonian. In the particular case of a chiral conformal theory,
this provides a local analogue of a global
formula obtained by Kac and Wakimoto within the  context of
representations of certain affine Lie algebras. Our formula is
model independent and its version in
general Quantum Field Theory applies to black hole
thermodynamics. The relative free
energy between two thermal equilibrium states associated with a
black hole turns out to be proportional to
the variation of the conditional entropy in different
superselection sectors, where the conditional  entropy is defined as
the  Connes-St\oe rmer  entropy associated with the DHR localized
endomorphism representing the sector. The constant of
proportionality is half of the Hawking temperature.
As a consequence the relative free energy is quantized
proportionally to the logarithm of a rational number, in particular
it is equal to a linear function the logarithm of an integer once
the initial state or the final state is taken fixed.
\vfill\eject

%%%%%%%% AUTO REF. %%%%%%%%%%%%%

\newcount\REFcount \REFcount=1
\def\numref{\number\REFcount}
\def\addref{\global\advance\REFcount by 1}
\def\wdef#1#2{\expandafter\xdef\csname#1\endcsname{#2}}
\def\wdch#1#2#3{\ifundef{#1#2}\wdef{#1#2}{#3}
    \else\write16{!!doubly defined#1,#2}\fi}
\def\wval#1{\csname#1\endcsname}
\def\ifundef#1{\expandafter\ifx\csname#1\endcsname\relax}
%%%%%%%%%%%%%%%%%%%%%%%%%%%%%%%%%%%%%%%%%%%%
\def\ref(#1){\wdef{q#1}{yes}\ifundef{r#1}$\diamondsuit$#1
  \write16{!!ref #1 was never defined!!}\else\wval{r#1}\fi}
\def\inputreferences{
    \def\REF(##1)##2\endREF{\wdch{r}{##1}{\numref}\addref}
    \REFERENCES}
\def\references{
    \def\REF(##1)##2\endREF{
        \ifundef{q##1}\write16{!!ref. [##1] was never quoted!!}\fi
        \item{[\ref(##1)]}##2}
    \par\REFERENCES}
%%%%%%%%%%%%%%%%%%%%%%%%%%
%%%%%% REFERENCES %%%%%%%%
%%%%%%%%%%%%%%%%%%%%%%%%%%
 \def\REFERENCES{
\REF(Ara){\smc Araki, H.} {\it Relative Hamiltonians for faithful
normal states of a von Neumann algebra,} Pub. R.I.M.S., Kyoto Univ.
{\bf 9} (1973), 165-209.\endREF
\REF(AHKT){\smc Araki H., Haag R.,
Kastler D., Takesaki M.,} {\it Extensions of KMS states and chemical
potential}, Commun. Math. Phys. {\bf 53} (1977), 97-134.
 \endREF
  \REF(BCW){\smc Bardeeen J.M., Carter B., Hawking S.W,} {\it The
four laws of black holes mechanics},  Commun. Math. Phys.
{\bf 31} (1973), 161. \endREF
 \REF(Bek){\smc Bekenstein, J. D.} {\it Black holes and entropy},
Phys. Rev.
{\bf D7} (1973), 2333. \endREF
 \REF(BCL){\smc Bertozzini P., Conti R., Longo, R.} {\it Covariant
 sectors with infinite dimension and positivity of the energy},
 Preprint.
 \endREF
 \REF(BW){\smc Bisognano J., Wichmann E.,} {\it On the duality
condition for a Hermitian scalar field}, J. Math. Phys. {\bf
16} (1975), 985-1007 and J. Math. Phys. {\bf
17} (1976), 303-321.
 \endREF
  \REF(BR) {\smc Bratteli O., Robinson D.W.,} {\it Operator
Algebras and Quantum Statistical Mechanics, II.} Sprin\-ger-Verlag,
Berlin-Heidelberg-New York, 1981.
 \endREF
 \REF(BGL) {\smc Brunetti R., Guido D., Longo R.,} {\it Modular
structure and duality in conformal quantum field theory},
Commun. Math. Phys. {\bf 156} (1993), 201--219.$\quad$\endREF
\REF(BGL2) {\smc Brunetti R., Guido D., Longo R.,} {\it Group
cohomology, modular theory and spacetime symmetries}, Rev.
Math. Phys. {\bf 7} (1995), 57--71.\endREF
\REF(BDL){\smc
Buchholz D. D'Antoni C., Longo R. }
   {\it Nuclear
maps and modular structures. I,} J. Funct.
Anal.  {\bf 88} (1990), 223-250,
II  Commun. Math. Phys. {\bf 129} (1990), 115--13. \endREF
\REF(BDopL){\smc
Buchholz D. Doplicher S., Longo R. }{\it On Noether's
theorem in quantum field theory}, Ann. Phys. {\bf 170} (1986),
1-17.\endREF
\REF(BF) {\smc Buchholz D. Fredenhagen K. } {\it Locality and
structure of particle states,}   Commun. Math. Phys.  {\bf  84}
(1982), 1-54.\endREF
 \REF(Con1) {\smc Connes A. } {\it Une
classification des facteurs de type III}, Ann. Sci. Ec. Norm. Sup.
{\bf 6} (1973), 133--252.
 \endREF
  \REF(Con0){\smc  Connes A.} {\it On a spatial theory of von
Neumann algebras,} J. Funct. An. {\bf 35} (1980), 153-164.\endREF
  \REF(Con2){\smc  Connes A.} {\it Entropie de Kolmogoroff-Sinai et
  m\'ecanique statistique quantique,} C. R. Acad. Sci. Paris S\'er.
I {\bf 301} (1985), 1-6.\endREF
 \REF(CS) {\smc Connes A., St\oe rmer, E.} {\it Entropy for
automorphisms
 of $II_1$ von Neumann algebras,} Acta Math. {\bf 134} (1975),
288-306.\endREF
\REF(Dav) {\smc Davies P.C.W.} {\it Scalar particle production in
Schwarzschild and Rindler metrics,} J. Phys. {\bf A8} (1975), 608.
\endREF
 \REF(DHR){\smc Doplicher S., Haag R., Roberts J.E.} {\it Local
observables and particle statistics I}, Commun. Math. Phys.
{\bf 23} (1971), 199-230, $II$ Commun. Math. Phys.
{\bf 35} (1974), 49-85.
 \endREF
 \REF(DR){\smc Doplicher S., Roberts J.E.} {\it Endomorphisms of
 C$^*$-algebras, crossed produts and duality for compact groups},
Ann. Math. {\bf 170}  (1989), 75.
 \endREF
\REF(EK){\smc Evans D., Kawahigashi K.} {\it Quantum Symmetries
on Operartor Algebras,} (in press).\endREF
\REF(FJ) {\smc Fredenhagen K., J\"or\ss\ M.} {\it Conformal
Haag-Kastler nets, pointlike localized fields and the existence of
operator product expansion}, Commun. Math. Phys. {\bf 176} (1996), 541.
 \endREF
\REF(FG) {\smc Fr\"ohlich J., Gabbiani F. }
{\it Operator algebras and conformal field theory,}
Commun. Math. Phys. {\bf 155} (1993),   569-640.\endREF
 \REF(GL1){\smc Guido D., Longo R.} {\it Relativistic invariance
and charge conjugation in quantum field theory}, Commun. Math.
Phys. {\bf 148} (1992), 521-551.
 \endREF
 \REF(GL2){\smc Guido D., Longo R.} {\it The conformal spin and
statistics theorem}, Commun. Math. {\bf 181} (1996), 11-35.
\endREF
   \REF(Haa){\smc Haag R.} {\it Local Quantum Physics}, Springer
Verlag, Berlin Heidelberg 1992.
 \endREF
 \REF(HHW){\smc Haag R., Hugenoltz N.M., Winnink M.} {\it On the
equilibrium states in quantum statistical mechanics}, Commun. Math.
Phys. {\bf 5} (1967), 215.\endREF
 \REF(HK){\smc Haag R., Kastler D.} {\it An
algebraic approach  to Quantum Field Theory}, J. Math. Phys. {\bf 5} (1964),
848-861.
 \endREF
 \REF(Haw){\smc Hawking S.W. }
 {\it Particle creation by black holes,}
Commun. Math. Phys. {\bf 43} (1975), 199.
 \endREF
 \REF(Hia){\smc Hiai, F.} {\it Minimum index for subfactors and
entropy. II,}  J. Math. Soc. Japan
{\bf 43 } (1991), 673--678
 \endREF
 \REF(HL){\smc Hislop P.D., Longo R.} {\it Modular
 structure of the
von Neumann algebras associated with the free scalar
massless field theory,}
 Commun. Math. Phys.    {\bf 84} (1982),   71-85.
 \endREF
 \REF(KW){\smc Kac V.G., Wakimoto M.} {\it Modular and conformal
invariance constraints in
representation theory of affine algebras}, Adv. in Math. {\bf 70},
 (1988), 156-236.\endREF
 \REF(KayWald){\smc Kay, B., Wald, R.} {\it Some recent
developments related to the Hawking effect}, In: Proc. Int.
Conf. Diff. Geom. Meth. in Ther. Phys., Doebner and Hennin eds..
World Scientific, Singapore 1987.\endREF
 \REF(Kos) {\smc Kosaki, H. } {\it Extension of Jones'
theory on index to arbitrary subfactors}, J. Funct. Anal. {\bf 66}
(1986), 123--140.\endREF \REF(Izu) {\smc Izumi,
M.} {\it Canonical extension of endomorphisms of factors}, in:
``Subfactors'', Proc. Taniguchi symposium, Araki, Kawahigashi,
Kosaki eds., World Scientific, Singapore 1994. \endREF
\REF(Jon)
{\smc Jones, V.R.F. } {\it Index for subfactors}, Invent. Math. {\bf
72} (1983), 1--25. \endREF
\REF(L0) {\smc Longo R.}, {\it Algebraic and modular structure of
von Neumann algebras of Physics}, Proc. Symp. Pure
Math. {\bf 38}, (1982), Part 2, 551.\endREF
 \REF(L1) {\smc Longo, R.} {\it Index of subfactors and statistics
of quantum fields. I }, Commun. Math. Phys. {\bf 126} (1989), 217--247.
 \endREF
 \REF(L2) {\smc Longo, R. } {\it Index of subfactors and statistics
of quantum fields. $II$ Correspondences, braid group statistics and
Jones polynomial}, Commun. Math. Phys. {\bf 130} (1990), 285--309.
\endREF
\REF(L3){\smc  Longo, R.} {\it Minimal index and braided
subfactors},  J. Funct. Anal. {\bf 109} (1992),
98-112. \endREF
\REF(L4) {\smc Longo R.} {\it Von~Neumann algebras and quantum
field theory,} Proceedings of the ICM, Z\"urich 1994, Birkh\"auser
Verlag, Basel, Switzerland 1995.
 \endREF
\REF(LR){\smc Longo R. Roberts, J.E.} {\it A theory of
dimension,}  K-Theory {\bf 11} (1997), 103-159.
 \endREF
 \REF(PP) {\smc Pimsner, M., Popa, M.,}  {\it Entropy and index for
subfactors}, Ann. Sci. Ec. Norm. Sup. {\bf 19} (1986),
57--106.\endREF
\REF(Reh) {\smc Rehren K.H. }
{\it On the range of the index of subfactors,}
J. Funct. Anal. {\bf 134} (1995) 183-193.\endREF
 \REF(Sch){\smc Schroer B.} {\it Some useful properties of rotational
 Gibbs states in chiral conformal QFT}, manuscript 1995.\endREF
 \REF(Sew1){\smc Sewell G.L.} {\it Quantum fields on manifolds:
PCT and gravitationally induced thermal states,}, Ann. Phys.
{\bf 141} (1982), 201.
 \endREF
 \REF(Sew2) {\smc Sewell G.L.} {\it On the generalised second low of
thermodynamics,} Phys. Lett. {\bf 122A} (1987), 309-311.
\endREF
  \REF(SV) {\smc Summers S., Verch R.} {\it Modular inclusion,
the Hawking temperature and Quantum Field Theory in curved
space-time,} Lett. Math. Phys. {\bf 37} (1996) 145-158.
\endREF
 \REF(SW) {\smc Streater R.F., Wightman A.S.} {\it PCT, Spin and
Statistics,  and all that}, Benjamin, Reading (MA) 1964.
\endREF
\REF(Tak) {\smc Takesaki M.} {\it Tomita theory
of modular Hilbert algebras,} Lect. Notes in Math. {\bf 128},
Springer Verlag, New York--Heidelberg--Berlin 1970.\endREF
\REF(Tak2) {\smc Takesaki M.} {\it Conditional expectation in von
Neumann algebras} J. Funct. Anal. {\bf 9} (1972), 306-321.\endREF
 \REF(Th) {\smc Thirring W.} {\it A Course in Mathematical Physics,}
Vol. 4, Springer Verlag, New York--Heidelberg--Berlin 1981.\endREF
 \REF(Unr) {\smc Unruh W.G.} {\it Notes on black hole
evaporation, } Phys. Rev.    {\bf D14} (1976), 870--892.\endREF
 \REF(W) {\smc Wald R. M.} {\it Quantum Field Theory in Curved
Spacetime and Black Hole Thermodynamics,} Univ. Chicago press
1994.\endREF
 \REF(WWW) {\smc Wick G.C., Wightman A.S., Wigner E.P.} {\it The
intrinsic parity of elementary particles}, Phys. Rev. {\bf 88}
(1952), 101-105. \quad\endREF
\REF(Wie) {\smc Wiesbrock, H. W.} {\it Superselection structure
and localized cocycles,} Rev. Math. Phys. {\bf 7} (1995),
127.\endREF}

\inputreferences

\NoBlackBoxes

\pagewidth{125mm}
\pageheight{195mm}
\parindent=8mm
\frenchspacing \tenpoint
\TagsOnRight
\line{}
\NoBlackBoxes

\def\en{\text{ End}(M)}
\def\proof{\demo\nofrills{\it Proof.\usualspace}}
\def\endproof{\hfill$\square$\smallskip\np}

\def\Co{{\Bbb C}}

\def\Re{{\Bbb R}}

%%%%%% SYMBOLS %%%%%%%%%%%%%
\def\A{{\Cal A}}

\def\a{\alpha}
\def\B{{\Cal B}}
\def\b{\beta}

\def\D{\Delta}

\def\E{{\Cal E}}
\def\f{\varphi}

\def\H{{\Cal H}}

\def\l{\lambda}
\def\L{\Lambda}
\def\m{\mu}

\def\o{\omega}
\def\O{{\Cal O}}
\def\p{\psi}

\def\Q{\Omega}
\def\r{\rho}

\def\s{\sigma}

\def\S{{\Cal S}}

\def\T{\Cal T}

\def\x{\xi}

\def\z{\zeta}

                     %Lorentz proper
           %Lorentz proper orthochronous
                     %Poincare\prime proper
\def\Ppo{{\Cal P}_+^\uparrow}           %Poincare\prime proper orthochronous

\def\Sp{\widetilde{\Cal P}_+}           %covering of Poincare\prime proper
\def\Spo{\widetilde{\Cal P}_+^\uparrow} %covering of Poincare\prime proper
orthochronous

\def\np{\par\noindent}
\def\lar{\longrightarrow}

\np{\fsection Introduction.} \smallskip\np
In the first part
of this paper we shall derive a formula
for the dimension of a superselection sector in Conformal Quantum
Field Theory.  However, as we shall see, the role played by
conformal invariance is not essential, and indeed we will
subsequently deal with general Quantum Field Theory and
apply our formula to the computation of the relative free energy
between two thermal equilibrium states of the background system for
a black hole. The reader  mainly interested in the latter topic may
at first read the second part of this introduction and then get
directly to Section 3.
\smallskip \np{\it A local analogue of the Kac-Wakimoto
formula.}  There is a general phenomenon relating the  distribution
of the Hamiltonian density levels to a dimension, a classical
example being given by  Weyl's theorem on the asymptotic distribution of the
Laplacian eigenvalues  on a compact Riemann manifold.

A similar occurrence appears in the context of  lowest
weight representations of certain affine Kac-Moody algebras. If
$L_0$ and $L_\r$  are the conformal Hamiltonians   in the
vacuum representation and in the
representation $\r$ of such an infinite dimensional Lie
algebra, then there exists the limit $$
\lim_{\b\to 0^+}{\text{Tr}(e^{-\b L_\r})\over \text{Tr}(e^{-\b
L_0})} = d(\r) \eqno (0.1)
$$
and the thus defined $d(\r)$ has the formal properties of a
dimension  [\ref(KW)].

One expects such a formula to hold in more generality in conformal
Quantum Field Theory on $S^1$ with $\r$ a superselection sector and
$d(\r)$ the statistical dimension of $\r$  (one has to
assume at least that $\text{Tr}(e^{-\b L_0})<\infty$ or any
structural property to garantee this), but no result in this
direction has been so far obtained (cf. [\ref(Sch)]).

However, if we restrict the vacuum state $\o$ to the local von
Neumann algebra $\A(I)$ associated with an interval $I$, then $\o$ is
faithful by the Reeh-Schlieder theorem and hence, by the
Tomita-Takesaki theory, $\o$ is a Gibbs state with respect to its
modular group\footnote{The reader should be aware of the different
meaning of the term {\it modular} in refs. [\ref(Tak)] and
[\ref(KW)].}. Such a modular group has a geometric meaning and we may
interpret it as a local dynamics, in other words  the logarithm of
the modular operator can be regarded as a local Hamiltonian. One
can then argue from [\ref(BDL)] that in this local situation the
distribution of the energy density levels should be tested in mean,
at a specific value of the inverse temperature parameter $\b$, and
hence also a local version of formula  (0.1) should not be
asymptotic as $\b\to 0^+$, but evaluated at a specific $\b$.

Let $\L_I(t)$ be the one-parameter group of special conformal
transformations of $S^1$ associated with the interval $I$ of $S^1$
(formula (2.1) below) and  $K_\r$ the generator of
 the corresponding one-parameter
unitary group in the representation $\r$.
We shall obtain the formula
$$
(e^{-2\pi K_\r}\x,\x)=d(\r)\eqno(0.2)
$$
where $\x$ is any cyclic vector for $\r(A(I'))$ such that
$(\r(\cdot)\,\x,\x)$ is the vacuum state on $\A(I')$ and
$I'=S^1\backslash I$.

In comparison with the formula (0.1), we  first note
that the right hand side $d(\r)$ in (0.2)  has not only the
properties of a dimension, but it is actually identified with the
Doplicher-Haag-Roberts [\ref(DHR)] statistical dimension of $\r$.

Moreover the local formula (0.2) holds in full generality,
independently of any requirement on the growth of Hamiltonian
spectral density. Hopefully it might lead to a model
independent proof of the formula (0.1), but it has its own interest.
As we shall see, its proof makes use of the knowledge of the
modular structure of the local von Neumann algebras
[\ref(BGL),\ref(FG)], the description of the conjugate sector in
terms of the modular involutions [\ref(GL1),\ref(GL2)], and a result
on actions of groups on tensor categories that determines $K_\r$ as
a linear function of the logarithm of the  relative modular operator
(as in formula (3.8) below).

Finally the validity of formula (0.2) goes beyond the context of
Conformal Quantum Field Theory. Indeed the same structure is present
in the general setting  Quantum Field  Theory on Minkowski
space [\ref(SW)], provided we consider the local von Neumann algebra
associated with a wedge region, because in this case the modular
structure has the geometric interpretation described by the
Bisognano-Wichmann theorem. We then treat this case, where a further
physical interpretation can be given. Mutatis mutandis, formulae
in Section 2 are also valid Section 3  and vice versa, with the
exception of Corollary 3.6. We  avoid repetitions and state in each
of them the results closer to the spirit of the section.

\smallskip\np {\it Quantum numbers for the relative free energy
associated with a black hole.} As was indicated  in
[\ref(BCW),\ref(Bek)], a black hole looks from the outside with some
features of a thermodynamical system in equilibrium. In
particular Bekenstein suggested the entropy  of a black hole to be
equal to $\lambda A$ where $A$ is the surface area of the black hole
and $\lambda$ a constant, an hypotesis related  the
Generalised Second Low of Thermodynamics $$
\text{d}S+\lambda\text{d}A\geq 0\eqno(0.3)
$$
where, in any process, $\text{d}A$ is the increment of $A$ and
$\text{d}S$ is the increment the entropy of the outside region.

Taking into account Quantum effects and General Relativity,
Hawking [\ref(Haw)] was led to the conclusion that the black
hole has a surface temperature
$$
T= {\hbar\over kc}{a\over 2\pi}
$$
where $a$ is the acceleration of a freely falling object at the
surface of the black hole\footnote{$T={c^3\hbar\over 8\pi kMG}$ in
terms of the mass $M$ of the star and the gravitational constant
$G$. In the following we shall always use natural units so that the
Plank costant $\hbar$, the speed of light $c$ and the Boltzmann
costant $k$ are all set equal to 1, thus $T=a/2\pi$.}. This
computation was made in the context of a free Quantum Field Theory
on a curved space-time.

Sewell [\ref(Sew1)] noticed that, at least in the case of a
spherically symmetric eternal black hole, a model independent
derivation of the
 Hawking temperature is possible, also in analogy with the Unruh
effect [\ref(Unr)], see also [\ref(Dav)], by means of the
Bisognano-Wichmann theorem [\ref(BW)] on the Minkowski space-time.
One considers the Rindler space-time $W$ as an approximation of the
Schwarzschild space-time and realizes $W$ as a wedge region in the
Minkowski space-time, say $W=\{x,\, x_1>|x_0|\}$. The evolution

$$
\left\{ \aligned
x_0(t) &=  a^{-1}\text{sh}t \\
x_1(t) &=  a^{-1}\text{ch}t
\endaligned \right.
$$
corresponds to an observer moving within $W$ with uniform
acceleration $a$, and his proper time is equal to $t/a$. $W$
is a natural horizon for this observer, since he cannot send a
signal out of $W$ and receive it back. The von Neumann algebra
$\A(W)$ of the observables on the Minkowski space localized in $W$
is therefore the proper (global) observable algebra for such a
mover. The
 proper time translations for him are thus given by the
one-parameter automorphism group $\a_{at}$ of $\A(W)$ corresponding
to the rescaled pure Lorentz transformations leaving $W$ invariant.
By the Bisognano-Wichmann theorem $\a_{at}$ satisfies the KMS condition at
inverse temperature
$$\b = {2\pi\over a}
$$
with respect to  the (restriction of
the Minkowski) vacuum state $\o$ to  $\A(W)$, i.e. the latter is a
thermal equilibrium state [\ref(HHW)]. By  Einstein equivalence
principle one can identify $W$ with the outside region of the black
hole and interpret the thermal outcome as a gravitational (black
hole) effect.

We refer to [\ref(Haa),\ref(KayWald),\ref(W)] for a more complete
account of these arguments and further references.
We note here that this description has certain restrictions. One is
due to the appearance of the Minkowskian vacuum tied up with the
Poincar\`e symmetries that do not exist globally on a general
curved space-time. Another point concerns the choice of the Rindler
space-time, that only near the horizon is a good
approximation of the more appropriate Schwarzschild space-time. We
shall briefly discuss these aspects in the final comments. Despite
its limitations, this viewpoint is
strikingly simple and powerful.

Let now consider a thermodynamical system $\Sigma$ placed in
the asympotically flat outside region of a black hole $B$.
The Hawking radiation creates a heat bath for $\Sigma$ and therefore
$\Sigma$ is an open system. Taking into account only observable
quantities, Sewell [\ref(Sew2)] inferred that the
right themodynamical potential for $\Sigma$ is the Gibbs free
energy,
rather than the entropy, and rederived the generalised second low
(0.3), where  the area term represents now a contribution to the
mechanical work done by $\Sigma$ on $B$.

Due to the Hawking effect, we have spontaneous creation of particles,
so that the system undergoes a change of quantum numbers. From the
Quantum Field Theory point of view, the system goes in a different
superselection sector [\ref(WWW)]. Following [\ref(HK),\ref(DHR)] we
thus consider a representation $\r$ of the quasi-local C$^*$-algebra
$\A = \cup \A(\O)^{-}$ that is localizable in any space-like cone
and has finite statistics. Under general conditions
$\r$ is (and we assume to be)  Poincar\`e covariant with positive
energy-momentum [\ref(GL1)]. We may assume that $\r$ is localized
within $W$ and  therefore the  restriction of $\r$ to  $\A(W)\cap\A$
has a normal extension to an endomorphism of its weak closure
$\A(W)$, that  will still be denoted by $\r_{|\A(W)}$.

The index-statistics theorem [\ref(L1),\ref(L4)] shows that the
map $$
\r\lar\r_{|\A(W)}
$$
is a faithful full functor of tensor categories, namely all the
information on the superselection structure  (charge transfers,
statistics,\dots) is visible within $\A(W)$, in particular
$$
d(\r) = \text{Ind}(\r)^{1\over 2}
$$
where $d(\r)$ is the DHR statistical dimension, i.e.
the order of the parastatistics, and $\text{Ind}(\r)$ is the
Jones index of $\r_{|\A(W)}$ (more precisely the minimal
index, see [\ref(Jon),\ref(L4)]).

In the sector $\r$ the proper time evolution is given by a
one-parameter automorphism group $\a_{at}^\r$ of $\A(W)$
corresponding to the pure Lorentz transformations leaving $W$
invariant,  $$
\a_t^\r\cdot\r =\r\cdot\a_t .
$$
As we shall see, $\a_{at}^\r$ admits a unique thermal equilibrium
normal state $\f_\r$ at the same inverse Hawking temperature $\b
={2\pi/a}$. If $K_\r$ is the generator of the unitary
implementation of the pure Lorentz in the $x_1$-direction, then
$$H_\r=aK_\r$$
is the proper Hamiltonian for our system $\Sigma$ in the sector
$\r$.

A similar structure appears in the
analysis of the chemical potential [\ref(AHKT)]. Adding one particle
is not a drastic change as our thermodynamical system has
essentially infinitely many particles, so we do not obtain an
inequivalent representation, namely $\r$ is normal on $\A(W)$.
There is however an important difference. The chemical potential
labels different equilibrium states at the same temperature for
the same dynamics, while we look at the
 $\f_\r$'s   as  equilibrium states with respect to
their own different dynamics  $\a_{at}^\r$, a fact
compatible with the
General Relativity context where the dynamics should
depend on the state as the matter has influence on the metric tensor.

Motivated by the above discussion, we will
consider the relative free energy between two thermal equilibrium
states for the system $\Sigma$ associated with the external region
of a black hole.   Guided by the thermodynamical expression $$
\text{d}F = \text{d}E - T\text{d}S
$$
we express the relative free energy of the states $\o$ and
$\f_\r$ by
$$
F(\o|\f_\r) = \f_\r(H_\r) - \b^{-1}S(\o|\f_\r)
$$
where $S$ is the Araki relative entropy [\ref(Ara)] of the two
states  and $\f_\r(H_\r)$ is the relative mean (internal) energy.

We shall find the relation
$$
F(\o|\f_\r) = -{1\over 2}\b^{-1}S_c(\r)\eqno(0.4)
$$
where $S_c(\r)$ is the Connes-St\oe rmer [\ref(CS)] conditional
entropy of $\r(\A(W))\subset\A(W)$.

Here we use the fact that, by the Pimsner-Popa theorem [\ref(PP)]
and the index-statistics theorem [\ref(L1)],  ${1\over
2}S_c(\r)$ equals the logarithm of the statistical dimension
$d(\r)$ of $\r$. As the latter takes only integral values by the DHR
theorem [\ref(DHR)], we conclude that the possible values for the
relative free energy are $$
F(\o|\f_\r) = -\b^{-1}\log(n),\quad n = 1, 2, 3,\dots
$$
namely the integer $n=d(\r)$ here acquaints the different
meaning of a quantum number labeling the relative free energy
levels.

The vacuum state $\o$ should play no specific physical role in
(0.4) and it is only a covenient reference state in our setting.
Indeed we will extend the formula to the case of two
arbitrary thermal equilibrium states $\f_\s$ and $\f_\r$
$$
F(\f_\s|\f_\r)={1\over 2}\b^{-1}(S_c(\s)-S_c(\r))\eqno(0.5)
$$
and therefore $F(\f_\s|\f_\r) = \b^{-1}\log({n\over m})$ is
$\b^{-1}$-times the logarithm of a rational number ${n\over m}$
where $n$ depends only on $\s$ and $m$ only on $\r$.

Finally we observe that formula (0.5) is consistent with the above
recalled interpretations of the increment $\text{d}A$ of surface area
of the black hole  [\ref(Bek),\ref(Sew2)] and, in a sense, it unifies
different points of views: the increments of the conditional
entropy of $\Sigma$, an information theoretical concept, is indeed
proportional to the increment of its free energy, a statistical
mechanics concept.

We summurize the conceptual scheme in the proof of our result in the
following diagram

$$
\CD
\text {Connes-St\oe rmer Entropy}    @>\text{Pimsner-Popa
Th.}>exp>        \text{Jones Index } \\ @V\text{Th. 3.4}V {1\over
2}\b^{-1}V				             @V\sqrt{\cdot}V \text{Index-statistics
Th.} V  \\  \text{Relative Free Energy}   @<\b^{-1}log<\text{Cor.
2.2}< \text{DHR dimension} \endCD
$$

\medskip\np
{\fsection 1. Connes cocycles and endomorphisms.}\smallskip\np
Let $M$ be a von
Neumann algebra and $\f$, $\o$ faithful normal positive linear
functionals on $M$ and denote by $\s^\o$ and $\s^\f$ their modular
group given by the Tomita-Takesaki theory [\ref(Tak)]. The Connes
Radon-Nikodym cocycle [\ref(Con1)]
$$
u(t)=(D\f:D\o)_t $$
is the map $(t\in\Re\to u(t)\,\text{ unitary \,of \,} M)$
such that
$$
u(t+s) = u(t)\s^\o_t(u(s))\eqno(1.1)
$$
characterized by: for any given $x,y\in M$ there
exists a complex function $F$ bounded and continuous in $\{0\leq
\text{ Im}z\leq 1\}$ and analytic in its interior such that
$$
F(t)=\f(\s_t^\f(y)u(t)x),\quad F(t+i)=\o
(xu(t)\s^\o_t(y)).\eqno(1.2)
$$
The relevant property is that $u(t)$ intertwines $\s_t^\o$ and
$\s_t^\f$, namely $$\s^\f_t=u(t)\s_t^\o(\cdot)u(t)^*.\eqno (1.3)$$
Conversely a continuous
unitary $\s^\o$-cocycle $u$ (i.e.  (1.1) is valid for $u$) is the
Connes cocycle with respect to a unique  faithful normal
positive linear functional or semifinite weight
$\f$ of $M$. If $M$ is a factor, a continuous
unitary $\s^\o$-cocycle $u$
satisfying (1.3) is uniquely determined up to a
one-dimensional character of $\Re$, hence, in order to check whether
$u(t)=(D\f:D\o)_t$, one may test
equation (1.2)  in the special case $x=y=1$:
 there must
exist a function $F$ bounded and continuous in $\{0\leq \text{
Im}z\leq 1\}$ and analytic in its interior such that
$$
F(t)=\f(u(t)),\quad F(t+i)=\o (u(t))\eqno(1.4)
$$
In particular, given the $\s^\o$-cocycle $u$, the positive functional
$\f$
such that $u(t)=(D\f:D\o)_t$ may be computed by (1.2) as
$$
\f(x)= {\underset{t\to
-i}\to {\text {anal.\, cont.\,}}}\o(xu(t))\eqno(1.5)
$$
Let now $M$ be
an infinite factor and denote by $\en$ the set {\it finite-index}
(or {\it finite-dimensional}) endomorphisms of $M$. Namely $\r\in
\en$ if $\r$ is an endomorphism of $M$ whose intrinsic dimension
$d(\r)$ is finite ([\ref(LR)], see Appendix A). Equivalently
$\r(M)$ is subfactor of $M$ with finite index in the sense of
[\ref(Jon),\ref(Kos)], indeed, as shown in [\ref(L2)],
$$
d(\r)=\text{Ind}(\r)^{1\over 2},
$$
where $\text{Ind}(\r)$ denotes the minimal index of
$\r(M)\subset M$. We refer to [\ref(L4)] and references
therein for the notions of index theory of
our use.

We fix a normal faithful state $\o$ of $M$.
Given $\r\in\en$ we denote by $\Phi_\r$ the {\it minimal left
inverse } of $\r$ and set
$$
\Psi_\r=d(\r)\Phi_\r,\quad\p_\r=\o\cdot\Psi_\r
$$
so that $\p_\r$ is a normal faithful positive functional of $M$.

We define the {\it cocycle of an endomorphism} $\r$ by
$$
u(\r,t)= (D\p_\r:D\o)_t\eqno(1.6)
$$
(see also [\ref(Izu)]).

As will be apparent, several of the results in this section
are valid (essentially with the same proofs) for endomorphisms of
factors with a normal faithful conditional expectation onto the
range, but for simplicity we just treat the finite-index case.
\proclaim\nofrills{\bf Proposition 1.1\usualspace} Let $\r\in\en$ and set
$\r_t=\s^\o_t\r\s^\o_{-t}$. Then $u(\r,t)$ satisfies
$$\text{Ad}u(\r,t)\cdot\r_t=\r\eqno(1.7)$$ In particular, if $\r $
is irreducible, then $u(\r,t)$ is characterized by (1.7) up to the
multiplication by a one-dimensional character of $\Re$. \endproclaim
\proof
The   minimal expectation $E_\r=\r\Phi_\r$ onto $\r(M)$ leaves
$\p_\r$ invariant since
$$\p_\r\r\Phi_\r=\o\Phi_\r\r\Phi_\r=\o\Phi_\r=\p_\r$$
thus
$${\s^{\p_\r}}_{|\r(M)}=\s^{\p_\r|\r(M)}=\s^{\o\r^{-1}}=
\r\s^\o\r^{-1},
$$
 namely
$$
\s_t^{\p_\r}\r(x)=\r\s^\o_t(x),\quad x\in M
$$
and since $\s_t^{\p_\r}=\text{ Ad}u(\r,t)\s^\o_t$ we have
$$
\text{ Ad}u(\r,t)\cdot\s^\o_t\cdot\r\cdot \s^\o_{-t}=\r.\eqno(1.8)
$$
Now the equation (1.8) detemines the restriction of
$\text{ Ad}u(\r,t)$ to $\r(M)$ hence it determines $u(\r,t)$
up to muliplication by a unitary in $\r(M)'\cap M$ and therefore,
if $\r$ is irreducible, it determines  $u(\r,t)$ up to a phase.
\endproof
As recalled in Appendix A, $\en$ are the objects of
a tensor C$^*$-category where the arrows are given by (A.1).
The following in this section has a natural
interpretation in the setting of tensor  C$^*$-categories,although
here below we use the explicit formulas of our setting (see
[\ref(L2)]).

As we shall use the greek letter $\s$ to denote an element of $\en$,
the modular group will be always denoted with explicit reference to
the functional (e.g. $\s_t^\o$).
\proclaim\nofrills{\bf Proposition 1.2\usualspace} Let
$\r\in\en$ be irreducible and contained in  $\s\in\en$, namely there
exists an isometry  $w$ in $(\r,\s)$ (i.e. $w\in M$ and
$w\r(x)=\s(x)w,\,\forall x\in M$), then  $$ \Psi_\r=\Psi_\s(w\cdot
w^*). \eqno (1.9) $$ If $\s=\oplus_{i=1}^N n_i\r_i$ is an
irreducible decomposition of $\s$ and for each $i$  $\{w^{(i)}_k,\
k=1,\dots n_i\}$ is an orthonormal basis of isometries in
$(\r_i,\s)$, then
 $$
 \Psi_\s=\sum_{i=1}^N\sum_{k=1}^{n_i}
 \Psi_{\r_i}(w^{(i)*}_k\cdot w^{(i)}_k).\eqno (1.10)
 $$
\endproclaim
\proof
 Let $\Psi '_\s$
be defined by the right hand side of (1.10). Then
$$
\Psi_\s '(\s(x))=\sum_{i=1}^N\sum_{k=1}^{n_i}
 \Psi_{\r_i}(w^{(i)*}_k \s(x)w^{(i)}_k) =
\sum_{i=1}^N\sum_{k=1}^{n_i}
 \Psi_{\r_i}(\r_i(x)) = \sum_{i=1}^N n_i d(\r_i)x = d(\s)x
 $$
 thus $d(\s)^{-1}\Psi_\s '$ is a left inverse of $\s$. As
 $\Psi'_\s(w^{(i)}_k w^{(i)*}_k)=\Psi_{\r_i}(1)=d(\r_i)$
 we have that $\Psi_\s ' = \Psi_\s$, showing  the
validity of the equation (1.10).  Formula (1.9) is obtained
similarly. \endproof
\proclaim\nofrills{\bf Proposition 1.3\usualspace}  If $T$ is an arrow in
  $(\r,\s)$, then
$$
Tu(\r,t)=u(\s,t)\s^\o_t(T)\, .\eqno(1.11)
$$
\endproclaim
\proof Let first $\r$ be an irreducible component of $\s$ and $w$
an isometry in $(\r,\s)$. Then by (1.9) we have
$$
u(\r,t)=(D\p_\r:D\o)_t=(D\p_\s(w\cdot w^*):D\o)_t
$$
therefore by the Radon-Nikodym chain rule and using the relation
$(D\p_\s(w^*\cdot w):D\p_\s)_t=w^*\s^{\p_\s}_t(w)$ we have
$$
u(\r,t)=(D\p_\s(w^*\cdot w):D\p_\s)_t(D\p_\s:D\o)_t
=w^*\s^{\p_\s}_t(w)u(\s,t)=w^*u(\s,t)\s^\o_t(w)
$$
and after multiplying on the left by $w$ all members in the above
expression, and using the $\s_t^{\psi_\s}$-invariance of $ww^*$, we
obtain the special case of (1.11)
$$
wu(\r,t)=u(\s,t)\s_t^\o(w).\eqno(1.12)
$$
If now $\s=\oplus_{i=1}^N n_i\r_i$ is an irreducible decomposition of
$\s$ and the $w_k^{(i)}$ are as in Proposition 1.2, we have
from (1.12)
$u(\s,t)w_k^{(i)}w_k^{(i)*}=w_k^{(i)}u(\r,t)\s^\o(w_k^{(i)*})$
and summing up over $i$ and $k$ we have
$$
u(\s,t)=\sum
_{i,k}w_k^{(i)}u(\r,t)\s^\o(w_k^{(i)*}).\eqno(1.13)
$$
If now $T$ is an arrow in $(\r,\s)$, we may decompose both $\s$ and
$\r$ into irreducibles so that the ranges of the $w^{(i)}_k$'s
of $\r$ (resp. of $\s$) are either orthogonal or contained in the
range of $T$ (resp. of $T^*$). Then multiplying (1.13) on the left by
$T$ (resp. on the right by $\s^\o_t(T)$) will kill the indeces
corresponding to the orthogonal part and the result is obtained by
the equation (1.12).
\endproof
\proclaim\nofrills{\bf Proposition
1.4\usualspace}$u(\r\s,t)=\r(u(\s,t))u(\r,t)$.
\endproclaim
\proof
We first note that
$$
\r(u(\r,t))=(D\p_{\r\s}:D\p_\r)_t\ .\eqno(1.14)
$$
Indeed as both functionals $\p_{\r\s}$ and $\p_\r$ leave invariant
the conditional expectation $\r\Phi_\r$ onto $\r(M)$, their
Radon-Nikodym cocycle coincides with the cocycle of their
restriction to $\r(M)$, thus
$$\align
{(D\p_{\r\s}:D\p_\r)}_t=
&{(D{\p_{\r\s}}_{|\r(M)}:D{\p_\r}_{|\r(M)})}_t\\
=&{(D\p_\s\cdot{\r^{-1}}_{|\r(M)}:D\o\cdot{\r^{-1}}_{|\r(M)})}_t\\
=&\r((D\p_\s:D\o)_t) = \r(u(\r,t)).
\endalign
$$
The Proposition then follows by the Radon-Nikodym chain rule
for the Connes cocycles.
\endproof
Propositions 1.3 and 1.4 state that $u(\r,t)$ is a
{\it two-variable-cocycle} (Appendix A), with respect to the action
of $\Bbb R$ on $\en$ given by
$$t\to \r_t =
\s_t^\o\cdot\r\cdot\s_{-t}^\o,\quad t\to\s_t^\o(T),\eqno(1.15)
$$
where $\r$ is an object and $T$ is an arrow in $\en$.

Recall now that each $\r\in\en$ has a conjugate object $\bar\r$,
namely there exist $R_\r \in
(\iota, \bar \rho \rho)$ and $\bar R_\r \in (\iota, \rho \bar \rho)$
standard solutions of the equation (A.1), i.e.
$$
\bar R_{\r}^*\r(R_{\r})=1, \quad
R_{\r}^*\bar\r(\bar R_{\r})=1
$$
and $\|R_{\r}\|=\|\bar
R_\r\|\, (=\sqrt{d(\r)})$ is minimal.

As explained in the Appendix A, given an arrow $T\in (\r_1,\r_2)$,
the conjugate arrow $T^{\bullet}\in (\bar\r_1,\bar\r_2)$ is  defined
by
$$
T^{\bullet}=\bar\r_2(\bar R_{\r_1}^*T^*)R_{\r_2}
$$
where $R_{\r_i}$ and $\bar R_{\r_i}$ give a standard solution for
the conjugate equation defining the conjugate $\bar\r_i$.

Note now that given $\r\in\en$, once we choose $\bar\r$ defined by
the $R_\r$ and $R_{\bar\r}$, than $(\bar\r)_t$ is a conjugate of
$\r_t$ defined by $R_{\r_t}=\s^\o_t(R_\r)$, $\bar
R_{\r_t}=\s^\o_t(\bar R_\r)$, that we may simply denote by
$\bar\r_t$. In the following $u(\r,t)^\bullet$ is defined by this
choice of the $R$-operators.
\proclaim\nofrills{\bf Proposition 1.5\usualspace}
$u(\bar\r,t)=u(\r,t)^\bullet$.
\endproclaim
\proof
By definition
$$\align
u(\r,t)^\bullet &= \bar\r(\bar R^*_{\r_t}u(\r,t)^*)R_\r\\
&= \bar\r(\s^\o_t(\bar R^*_\r)u(\r,t)^*)R_\r\\
&= \bar\r(\s^\o_t(\bar R^*_\r)\s^\o_t(u(\r,-t)))R_\r\\
&= \bar\r(\s^\o_t(\bar R^*_\r u(\r,-t)))R_\r\\
&= u(\bar\r,t)\s^\o_t(\bar\r(\bar R_\r^*
u(\r,-t)))u(\bar\r,t)^*R_\r\\
&= u(\bar\r,t)\s^\o_t(\bar\r(\bar R_\r^*
u(\r,-t)))\s^\o_t(u(\bar\r,-t))R_\r \\
&= u(\bar\r,t)\s^\o_t(\bar\r(\bar
R_\r^* u(\r,-t))u(\bar\r,-t))R_\r\endalign
$$
Thus we have to show that
$\s^\o_t(\bar\r(\bar
R_\r^* u(\r,-t))u(\bar\r,-t))R_\r=1$ or, by applying
$\s^\o_{-t}$, that  $\bar\r(\bar R_\r^*
u(\r,-t))u(\bar\r,-t)R_{\r_{-t}}=1$.
Indeed we have
$$\align
\bar\r(\bar R_\r^* u(\r,-t))u(\bar\r,-t)R_{\r_{-t}} &=
\bar\r(\bar R_\r^*)\bar\r(u(\r,-t))u(\bar\r,-t)R_{\r_{-t}}\\ &=
\bar\r(\bar R_\r^*)u(\bar\r\r,-t)R_{\r_{-t}}=
\bar\r(\bar R_\r^*)R_\r = 1\, ,\endalign
$$
where $u(\bar\r\r,-t)R_{\r_{-t}} = R_\r$ by Proposition 1.3.
\endproof
\proclaim\nofrills{\bf Lemma 1.6\usualspace} If $j$ is an $\o$-preserving
anti-automorphism  of $M$,
then
$$
u(\r,t)= j(u(j\cdot\r\cdot j^{-1},-t)).
$$
\endproclaim
\proof Since $j$ preseves $\o$, by the KMS condition
 $j\s_t^\o j^{-1}=\s_{-t}^\o$, thus \linebreak $j(u(j\cdot\r\cdot
j^{-1},-t))$ is  a $\s_t^\o$-cocycle and it coincides with
$u(\r,t)$ because it satifies the equation (1.5).
\endproof
\proclaim\nofrills{\bf Proposition 1.7\usualspace} Let $\T$ be a $C^*$-tensor
subcategory with conjugates of $\en$ and $z$ a two-variable cocycle
for the action $\Re\to\text{Aut}\T$ given by the modular group
$\s^\o$ (equation (1.15)). Suppose there is an anti-automorphism $j$
of $M$ such that $j\r j^{-1}$ is a conjugate of $\r$  and
$z(j\r j^{-1},t)=j(z(\r,-t))$ for a given $\r\in\T$. Then $z(\r,t)$
coincides with the Connes cocycle $u(\r,t)$ defined by (1.5). As a
consequence $$ d(\r)={\underset{t\to
-i}\to {\text {anal.\, cont.\,}}}\o(z(\r,t)).\eqno(1.16)
$$
\endproclaim
\proof We have $z(\r,t)=\mu(\r,t)u(\r,t)$ for some
character $\mu(\r,\cdot)$ of $\Re$. Since $\bar\r=j\r j^{-1}$ is a
conjugate of $\r$ we have by the Lemma 1.6
$$
z(\bar\r,t)=j(z(\r,-t))=j(\mu(\r,-t)u(\r,-t))
=\mu(\r,t)j(u(\r,-t))=\mu(\r,t)u(\bar\r,t)
$$
On the other hand by Lemma A.3 of the Appendix A
$$
z(\bar\r,t)=z(\r,t)^\bullet=\mu(\r,t)^\bullet u(\r,t)^\bullet
= \overline{\mu(\r,t)}u(\bar\r,t)
$$
thus $\mu(\r,t)=1$. Formula (1.16) is thus a consequence of (1.5)
in the case $x=1$.
\par\endproof
Before concluding this section we recall the notions of entropy of
later use.
If $N\subset M$ is an inclusion of $II_1$-factors, the
Connes-St\oe rmer (conditional) entropy $H(M|N)$ is defined in
[\ref(CS)]. By the Pimsner-Popa Theorem  [\ref(PP)]
$$
H(M|N) = \log[M:N] \eqno (1.17)
$$
where $[M:N]$ is the Jones index of $N\subset M$ (if $N\subset M$
is irreducible or extremal). $H(M|N)$ is  generalized  to the type
$III$ setting in [\ref(Con2)]. If $\f$ is a normal faithful state of
$M$ one defines $$
H_\f(M|N) = \sup_{(\f_i)}\{\sum_i S(\f|\f_i) -
S(\f{_|}_N|\f_i{_|}_N)\} $$
where ${(\f_i)}$ varies among the sets of finitely many normal
positive linear  functionals $\f_1, \f_2,\dots \f_n$ of $M$ such
that $\sum_{i=1}^n \f_i =\f$, and $S(\cdot|\cdot)$ denotes the
Araki relative entropy between states, see [\ref(BR)] and
eq. (3.7) below.

If $E$ is a normal conditional expectation of $M$ onto $N$
one sets
$$
H_E(M|N)= \sup\{H_\f(M|N),\, \f\cdot E = \f\}.
$$
If moreover $N$ is a $III_1$-factor, then
$$
H_E(M|N)=H_\f(M|N)
$$
for any normal faithful state $\f$ such that $\f\cdot E = \f$
[\ref(Hia)]. $H_E(M|N)$ depends on the choice of a normal conditional
expectation $E$, but we simply write
$$
H(M|N)=H_{E_{\text{min}}}(M|N)
$$
where $E_{\text{min}}$ is the minimal conditional expectation.
The Pimsner-Popa equality (1.17) holds true without restrictions,
provided $[M:N]$ denotes the minimum index, see [\ref(Hia)].
Finally, if $\r$ is an endomorphism of $M$, we consider the
conditional entropy of $\r$
$$ S_c(\r)=H(M|\r(M)).\eqno(1.18)
$$
\medskip \np{\fsection 2. The formula in conformal field theory.}
\smallskip\np
{\it 2.1 Finite index case.} We now consider a precosheaf (net) $\A$
of  von Neumann algebras associated with a chiral conformal
quantum field theory. Namely $\A$ is a map
$$
I\lar \A(I)
$$
from the (proper) intervals of $S^1$ to von Neumann  algebras $\A(I)$
on a fixed Hilbert space $\H$ that satisfies:
\smallskip
 {\it Isotony}: if $I\subset \tilde I$ then $\A(I)\subset
\A(\tilde I)$,
\smallskip
{\it Locality}: $\A(I)$ and $\A(I')$ commute elementwise, where
$I'=S^1\backslash I$,
\smallskip
{\it M\"obius covariance with positive energy}: there exists a
unitary reperesentation $U$ of
M\"obius group $SL(2,\Re)/\{1,-1\}$, that for convenience we regard
as a representation of its universal covering group $G$, such that
$U(g)\A(I)U(g)^* = \A(gI)$ and the generator of the one-parameter
rotation subgroup is positive. We set
$$
\a_g(X) = U(g)XU(g)^*
$$
where $X$ is a {\it local operator}, namely $X$ belongs to some
$\A(I)$, \smallskip
{\it Existence and uniqueness of the vacuum}: there exists a unique
(up to a phase) $U$-invariant unit vector $\Q\in\H$, and it is cyclic for the
algebra generated by of local operators. We denote by
$\o=(\,\cdot\Q,\Q)$ the vacuum state.
\smallskip\np
For a discussion of these properties and their consequences, see
[\ref(GL2)].

Let $\r$ be a covariant positive energy {\it representation} of $\A$
on a separable Hilbert space $\H_\r$, namely for every proper
interval $I$ we have a representation $\r_I$ of $\A(I)$ on $\H_\r$
so that $$
{\r_{\tilde I}}_{|\A(I)} = \r_I \quad \text {if} \,\, I\subset \tilde
I $$
and a positive energy representation $U_\r$ of $G$ on $\H_\r$ such
that
$$
\r_{gI}(\a_g(X)) = U_\r(g)\r_I(X)U_\r(g)^*,
$$
$X\in\A(I)$ (if $\A$ is strongly
additive the covariance property is automatic [\ref(GL1)]).
We shall always assume the representations to be covariant with
positive energy.

Let $\L_I(t)$ be the special conformal one-parameter
group associated with $I$. If $I$ is the upper semi-circle
then
$$
\L_I(t)z={(z+1)-e^{-t}(z-1)\over
(z+1)+e^{-t}(z-1)}\eqno(2.1)
$$
while if $I_0$   is any other interval $\L_{I_0}$ is well defined
by  conjugation of the transformation (2.1) with a conformal
transformation $g\in SL(2,\Bbb R)$ such that $gI = I_0$. We denote
by
$$
K^I_\r = -i{\text{d}\over\text{d}t}U_\r(\L_I(t)){_|}_{t=0}
$$
the
infinitesimal generator of  $U_\r(\L_I(t))$.
\proclaim\nofrills{\bf Theorem 2.1\usualspace} Let $\r$ be a representation of
$\A$, $I\in S^1$ an interval and $\x\in\H_\r$  a  cyclic vector for
$\r_{I'}(\A(I'))$ such that $\o(X)=(\r_{I'}(X)\x,\x)$ for all
$X\in\A(I')$. Then\footnote{We shall use the following convention:
is $A$ is a positive selfadjoint operator and $\eta$ a vector, then
$(A\eta,\eta)=\|A^{1\over 2}\eta\|^2$ if $\eta$ belongs to the
domain of $A^{1\over 2}$, and  $(A\eta,\eta)=+\infty$ otherwise.}
$$
(e^{-2\pi K^I_\r}\x,\x)=d(\r). $$
A vector $\x$ as above always exists.
\endproclaim
\np As we are interested in the representation $\r$  up to unitary
equivalence, we may identify $\H_\r$ with $\H$ so that, due to Haag
duality (see below), $\r$ becomes an endomorphism of $\A$ localized
in a given interval $I$,  namely $\r_{I'}$ acts  identically, and
$\r_{\tilde I}(\A(\tilde I))\subset \A(\tilde I)$ if $I\subset
\tilde I$, see [\ref(GL2)]. Because of the Reeh-Schlieder
theorem (see [\ref(FJ)]), the vacuum vector $\Q$ is cyclic for any
local von Neumann algebra, in particular for $\A(I')$, therefore $\Q$
satisfies in this representation the properties required for the
vector $\xi$ in the statement of Theorem 2.1, showing its existence.

As $\r$ is covariant, there is a unitary $\a$-cocycle $z(\r,g)$
such that
$$
\text{ Ad}z(\r,g)\cdot\a_g\cdot\r\cdot\a_g^{-1}=\r, \eqno (2.2)
$$
where $\a_g=\text{Ad}U(g)$.
More precisely the equation $z(\r,g)=U_\r(g)U(g)^*$ defines
$z(\r,g)$ if $g$ belongs to a neighbourood of the identity of $G$,
and $z(\r,g)$ is localized in the sense that it belongs to
$\A(\tilde I)$ is $\tilde I$ is any interval  containing both $I$
and $gI$ [\ref(GL1)], (see also [\ref(Wie)]). Then $z(\r,g)$ is
defined for arbitrary $g\in G$ as an element of the universal
algebra $C^*(\A)$ by the cocycle identity
$z(\r,gh)=z(\r,g)\r(z(\r,h))$, but we do not need this fact.

It is important to note that if $\r$ is irreducible or a finite
direct sum of irreducibles, as is the case of a finite-index $\r$,
then $z$ is uniquely determined by the formula (2.2) as a localized
cocycle, because $G$ has no non-trivial unitary finite-dimensional
representation, see [\ref(GL2)].

We  consider the tensor $C^*$-category $\E_I$ whose objects are the
(covariant, positive energy) endomorphisms of $\A$ with finite
index,
namely
$$
d(\r)=d(\r_I) <\infty ,
$$
localized in an interval whose closure is contained
in the interior of a given interval $I$ and the arrows $(\r,\s)$ are
the local operators $T$ such that $T\r_{I_1}(X)=\s_{I_1}(X)T$ for
all intervals $I_1$ and local operators $X\in \A(I_1)$. By
[\ref(L1)] (see also [\ref(GL2)]) the restriction map
$$
\r\in\E_I\to\r_I\in\text{End}\A(I)
$$
is a faithful full functor, therefore we may identify $\E_I$ with a
tensor $C^*$-subcategory of End$\A(I)$, so that $d(\r)$ is
identified with the DHR statistical dimension of $\r$.

Then $z$ is in a natural
sense a {\it local two-variable cocycle} for the local action of $G$
on $\E_I$ given by $\r\to\a_g\r\a_g^{-1}$ ($\r\in\E_I$) and $T\to
\a_g(T)$ ($T$ arrow), namely the properties $a)$ and $b)$ defining a
two-variable cocycle after Lemma A.3 of the Appendix A hold for
$z(\r,g)$, but only if $g$ belongs to a neighbourhood of the
identity of $G$ (depending on $\r$). For example if $\r,\s\in\E_I$
then $\r(z(\r,g))z(\s,g)$ exists if $g$ lies in a neighbourhood of
the identity of $G$ and satisfies  (2.2) for $\r\s$, hence it agrees
with $z(\r\s,g)$ as the $\a$-cocycle property and formula (2.2)
determines it. We thus have: \proclaim\nofrills{\bf Lemma 2.2\usualspace}
$z(\r,g)$ is a local two-variable cocycle for the action of $G$ on
the tensor $C^*$-category $\E_I$.
\endproclaim
\proof
This follows by the above discussion and an elementary direct
verification of the two-variable cocycle property.\endproof
Recall now that each localized endomorphism $\r$ has a conjugate
localized endomorphism given by the formula [\ref(GL1)]
$$
\bar\r = j\cdot\r\cdot j\eqno(2.3)
$$
where $j = \text{Ad}J$ is the anti-automorphism of $\A$ implemented
by the modular conjugation $J$ of $(\A(I_1),\Q)$ for any choice of
the interval $I_1$. To be definite let  $I$ be the upper half-circle
and $I_1$ the right half-circle. Due to the geometrical meaning of
$J$, $j$ implements an anti-automorphism of $\A(I)$.
If $g$ is in the M\"obius group, we denote by $g^j$ its conjugate
by the anti-automorphism given by the reflection on $S^1$
corresponding to $j$.
\proclaim\nofrills{\bf Proposition 2.3\usualspace} Let $\r$ be finite-index
endomorphism of $\A$ localized in the interval $I$. With the above
notations, we have  $$
z({\bar\r},g) = z(\r,g)^\bullet = j(z(\r,g^j))
$$
(see Section 1 and Appendix A for the definition of the
$^\bullet$-mapping).
\endproclaim
\proof The first equality follows from Lemma 2.1 and Lemma A.3.
However one may see directly the validity of both equalities by the
uniqueness of $z({\bar\r},g)$ by checking that also $z(\r,g)^\bullet$
and $j(z(\r,g^j))$ are  local $\a$-cocycles and both satisfy
formula (2.2) for $\bar\r=j\r j$. \endproof
Now the modular structure of $\A$ is computed in
[\ref(HL),\ref(BGL)], in particular we have
$$
\Delta^{it}_I = U(\L_I(-2\pi t))
$$
where $\Delta_I$ is the modular operator of $(\A(I),\Q)$.
An important consequence is Haag duality
$$
\A(I')=\A(I)',
$$
moreover the $\A(I)$'s are type $III_1$ factors.

Next theorem computes the modular structure of $\A$ in the
representation $\r$, see also Proposition 3.5. We set
$u_I(\r,t)=u(\r_I,t)=(D\o{\Psi_{\r_I}}:D\o_{|\A(I)})_t$
as in (1.6), with $\o$ the vacuum state.
\proclaim\nofrills{\bf Theorem 2.4\usualspace} Let $\r$ be finite-index
endomorphism of $\A$ localized in the interval $I$. Then
$$
u_I(\r,t)=z(\r,\L_I(-2\pi t)),\ t\in\Re .
$$ \endproclaim \proof
By the formula (2.3) for the conjugate sector and  Lemma 2.2,
the theorem follows immediatly by Proposition 1.7.
\endproof
\proclaim\nofrills{\bf Corollary 2.5\usualspace} Let $\r$ be an irreducible
finite-index endomorphism of $\A$ localized in the interval $I$ of
$S^1$. Then  $$
d(\r)=(e^{-2\pi K^I_\r}\Q,\Q)=\|e^{-\pi
K^I_\r}\Q\|^2.\eqno(2.4)
$$
\endproclaim
\proof
Since $z_I(\r,\L_I(t)) = U_\r(\L_I(t))U(\L_I(t))^*$, Proposition
1.7 and Theorem 2.3 show that the function $t\to\o(z(\r,\L_I(-2\pi
t)))$  extends to a function bounded and continuous in the strip
$\{-1\leq\text{Im}z\leq 0\}$ and analytic in its interior so that
$$
d(\r)={\underset{t\to
-i}\to {\text {anal.\, cont.\,}}}\o(z(\r,\L_I(-2\pi t)))
= {\underset{t\to
-i}\to {\text {anal.\, cont.\,}}}(U_\r(\L_I(-2\pi t))\Q,\Q).
$$
Standard functional analysis arguments then show that the right
hand side of the above expression is equal to
$(e^{-2\pi K^I_\r}\Q,\Q)$. \endproof
\smallskip\np
{\it Proof of Theorem 2.1 in the finite index case}.  Let
$V:\H_\r\to\H$ be the unitary given by
$$
V\r(X)\x=X\Q,\quad X\in\A(I')
$$
so that $V\r(X)V^*=X$ if $X\in\A(I')$, thus
$$
\r'=V\r(\cdot)V^*
$$
is an endomorphism of $\A$ localized in $I$. By Corollary 2.5
we then have
$$
d(\r)=(e^{-2\pi K^I_{\r'}}\Q,\Q)=(Ve^{-2\pi K^I_\r}V^*\Q,\Q)
=(e^{-2\pi K^I_\r}\x,\x),
$$
in case $\r$ has finite index. We have already commented on the
existence of $\xi$. The infinite index case is discussed here
below.\endproof \smallskip\np
{\it 2.2 General case: a criterium for finite index.}
We now show
that formula (2.4) gives a criterium for the finiteness
of the index of a sector, namely that Theorem 2.1 holds
without restrictions. We
start with a general fact.
\proclaim\nofrills{\bf Proposition 2.6\usualspace} Let
$N\subset M$ be an inclusion of factors and assume that
there exist a normal faithful conditional expectation $E$ of $M$
onto $N$ and a normal faithful conditional expectation $E'$ of $N'$
onto $M'$.
Then  $N'\cap M$ is a discrete type I von
Neumann algebra, i.e. a (possibly infinite) direct sum of type I
factors. Moreover for each minimal projection $p$ of $N'\cap M$
the inclusion $Np\subset pMp$ has finite index.
\endproclaim
\proof Setting $R=N'\cap M$, $E$ restricts to a faithful
expectation of $N\vee R$ onto $N$, hence $N\vee R$ is canonically
isomorphic to the von Neumann tensor product $N\otimes R$ and we
can assume this isomorphism to be spatial (by tensoring
$N$ and $M$ by a type $III$ factor, if necessary). On the other
hand $E'$ factors through a faithful normal expectation of $N'$ onto
$(N\vee R)'= N'\cap R'$ by Takesaki's theorem [\ref(Tak2)], hence,
with the above identification, we have a normal faithful expectation
of $N'\otimes B(\H)$ onto $N'\otimes R'$, that restricts to a normal
expectation of $\Bbb C\otimes B(\H)$ onto $\Bbb C\otimes R'$, that
implies $R$ to be be a type $I$ von Neumann algebra.

As $R$ is a direct sum of homogeneous type $I_n$ von Neumann
algebras, by considering the reduced inclusion corresponding
to an abelian projection of $R$ (fixed by the modular group of the
expectation) we are left to prove our statement in the case $R$ be
an abelian von Neumann algebra, namely we have to prove that $R$ is
totally atomic, for in this case the finiteness of the index of
the reduced inclusions corresponding to minimal projections of $R$
would follow by [\ref(L2), Proposition 4.4].

By decomposing $R$ into its diffuse and atomic part, we may then
assume $R$ to be diffuse abelian and find an
absurd. To this end, for notational convenience, we may identify $N$
with $M$ (see [\ref(L3)]), i.e. we set $N=\s(M)$ for some
endomorphism $\s$ of $M$. We may decompose
$\s=\int^{\oplus}\s_\l\text {d}\m(\l)$ into irreducibles and as $R$
is abelian $\s_\l$  is disjoint to $\s_{\l'}$ for almost all pairs
$(\l,\l')$, hence $\bar\s_{\l}\s_{\l'}$ does not contain the
identity, execept for $(\l,\l')$ in a set of product measure 0 and
we conclude that $\bar\s\s$ does not contain the identity too. By
[\ref(L1)] this shows that there exists no normal faithful
expectation onto $N$ contradicting our hypotesis.  \endproof
\proclaim\nofrills{\bf Lemma 2.7\usualspace} Let $N\subset M$ be an
inclusion of von Neumann algebras on a Hilbert space $\H$, $\Q$ a
cyclic separating vector for $M$ and $\D_M$, $\D_N$ the
corresponding modular operators on $\H$ and on $\overline {N\Q}$.
Then $
\|\D_M^{1\over 2}\xi\| = \|\D_N^{1\over 2}\xi\|
$
for all $\xi$ in the domain of $\D_N^{1\over 2}$.
\endproclaim
\proof
If $x\in N$ we have with usual notations
$$
\|\D_N^{1\over 2}x\Q\| =\|J_N\D_N^{1\over 2}x\Q\|=\|x^*\Q\|
=\|J_M\D_M^{1\over 2}x\Q\|=\|\D_M^{1\over 2}x\Q\|
$$
and as $N\Q$ is a core for $\D_N^{1\over 2}$ the equality holds
for all the vectors in the domain of $\D_N^{1\over 2}$.
\endproof
\proclaim\nofrills{\bf Corollary 2.8\usualspace} Let $T(t)$ and $V(s)$ be two
unitary one-parameter groups on a Hilbert space $\H$ such that
$$
V(s)T(t)V(-s)=T(e^{-2\pi s}t),\ t,s\in\Bbb R\eqno(2.5)
$$
and assume $-i{\text{d}\over\text{d}t}T(t)_{|t=0}$ to be positive. Then
$$
\|e^{-\pi D}\xi\| = \|e^{-\pi D}T(t)\xi\|\eqno(2.6)
$$
for all $\xi$ in the domain of $e^{-\pi D}$ and all $t\geq
0$,
where $D$ is the generator of $V$.
\endproclaim
\proof The projection $P$ onto the $T$-fixed vectors commutes
both with $T$ and $V$ and on such vectors the equation (2.6)
trivially holds, hence we may assume that $P=0$. With this
assumption all non-zero representations of the commutation relation
(2.5) are quasi-equivalent by von Neumann uniqueness theorem
($D$ and the logarithm of the generator of $T(t)$ satisfy
the Heisenberg commutation relations),
hence we may verify the equation (2.6) in any given representation.

Let $\B$ be the conformal net on
$\Bbb R = S^1\backslash\{-1\}$ given
by the current algebra (see e.g. [\ref(EK)]), $T$ and $V$ the
translation and dilation unitary groups. Then $e^{-2\pi D}=\D_M$,
where $M=\B(0,+\infty)$ and $T(t)e^{-2\pi D}T(-t)=\D_N$, where
$N=\B(t,+\infty)$.
Lemma 2.7 then applies to the modular operators $\D_M$ and $\D_N$
with respect to the vacuum and gives equation (2.6).
\endproof
We shall now denote by $T^I(t)$ the one-parameter unitary
group of translations associated with $I$, namely
cutting $S^1$ and
identifing it with $\Bbb R$ so that $I$ is identified with $\Bbb
R^+$,  then $T^I(t)$ correspond to the translations on $\Bbb R$.
If $\r$ is an endomorphism of $\A$
localized in $I$, we shall then denote by $T^I_\r$ the corresponding
one-parameter unitary group in the representation $\r$.
\proclaim\nofrills{\bf Corollary 2.9\usualspace} Let
$\r$ be an endomorphism of $\A$ localized in the interval $I$. Then
$$ \|e^{-\pi K^I_\r}\xi\|=\|e^{-\pi K^I_\r}T^I_\r(t)\xi\| $$
for all $\xi$ in the domain of $e^{-\pi K^I_\r}$ and $t\geq 0$ .
\endproclaim
\proof
Immediate by Corollary 2.8.
\endproof
\proclaim\nofrills{\bf Proposition 2.10\usualspace} Let $\r$ be an
endomorphism of
$\A$ localized in the interval $I$. If $(e^{-2\pi
K^I_\r}\Q,\Q)<\infty$, then the formula
$$
\psi_\r(XY^*)=(e^{-\pi K^I_\r}X\Q, e^{-\pi K^I_\r}Y\Q)
$$
determines a positive normal linear functional $\psi_\r$ on $\A(I)$
such that \linebreak
$(D\psi_\r:D\o_{|\A(I)})_t =z(\r,\L_I(-2\pi t))$. \endproclaim
\proof
By Connes' theorem [\ref(Con1)] there exists a unique normal
faithful semifinite weight $\psi_\r$ on $\A(I)$ such that
$(D\psi_\r:D\o_I)_t = z(\r,\L_I(-2\pi t))$, where we shorten
notations like $\o_{|\A(I)} = \o_I$. Moreover $$
\D(\o_{I'}|\psi_\r)^{it}=z(\r,\L_I(-2\pi t))
\D(\o_{I'}|\o_I)^{it}
$$
where $\D(\cdot|\cdot)$ denotes the
Connes spatial derivative [\ref(Con0)] (between a weight on a von
Neumann algebra and a weight on its commutant), and since by (2.2)
$$ \D^{it}(\o_{I'}|\o_I)=\D_I^{it}=e^{-2\pi itK^I}
$$
we have
$$
e^{-2\pi K^I_\r} = \D(\o_{I'}|\psi_\r).\eqno(2.7)
$$
By assumptions $\Q$ thus belongs to the domain
of $\D(\o_{I'}|\psi_\r)^{1\over 2}$ and this implies that
$\psi_\r(1)=\|\D(\o_{I'}|\psi_\r)^{1\over 2}\Q\|^2<+\infty$,
namely $\psi_\r$ is everywhere defined. The Proposition now
readly follows by formula (1.4):
$$\align\psi_\r(XY^*) &= {\underset{t\to
-i}\to {\text {anal.\, cont.\,}}}\o(Y^* u(\r,t)\s^\o_t(X)) \\ &=
{\underset{t\to
-i}\to {\text {anal.\, cont.\,}}}
\o(Y^* z(\r,\L_I(-2\pi t))\s^\o_t(X))
= {\underset{t\to -i}\to {\text{anal.\, cont.\,}}}
(e^{-i2\pi tK_\r}X\Q,Y\Q).\endalign $$
Alternatively one could use
directly the expression given by Proposition 3.5. \endproof

\proclaim\nofrills{\bf Proposition 2.11\usualspace} Let
$\r$ and $\psi_\r$ be as in Proposition 2.10 and identify $I$ with
$\Bbb R^+$ as above.
Then  $$
\psi_\r(U_\r(g)XU_\r(g)^*)=\psi_\r(X), \quad X\in\A(I),\eqno(2.8)
$$
 provided $g\in G$ is a dilation
or a positive translation associated with $I$.

As a consequence there exists a positive linear functional
$\tilde\psi_\r$ on $\A_\z=\cup_{\ell}\A(\ell,+\infty)^{-}$, normal
on any $\A(\ell,+\infty)$, translation and dilation invariant
in the representation $\r$, defined by $$
\tilde\psi_\r(X)=\psi_\r (T^I_\r(t)XT^I_\r(t)^*)
$$
where $X\in \A(\ell,+\infty)$ and $t+\ell>0$. \endproclaim
\proof The second assertion clearly follows from the first one.
Formula (2.8) holds if $g$ is a dilation, as the
dilations correspond to the  modular automorphisms of $\A(I)$ with
respect to $\psi_\r$, due to the construction of $\psi_\r$.
By Proposition 2.10 we thus have to show that
$$
\|e^{-\pi K_\r^I}T^I_\r(t)XT^I_\r(-t)\Q\|=
\|e^{-\pi K_\r^I}X\Q\|,\eqno(2.9)
$$
where $X\in\A(I)$ and $t\geq 0$.
Indeed by Corollary 2.9  for all $X\in\A(I)$ and $t\geq 0$ we have
$$\aligned
\|e^{-\pi K_\r^I}T^I_\r(t)XT^I_\r(-t)\Q\| &=
\|e^{-\pi K_\r^I}XT^I_\r(-t)\Q\| \\
&=\|e^{-\pi K_\r^I}XT^I_\r(-t)\Q\|=
\|e^{-\pi K_\r^I}X\xi\|. \endaligned\eqno(2.10)
$$
where $\xi=T^I_\r(-t)\Q=T^I_\r(-t)T^I(t)\Q$.
On the other hand  $z(\r,t)=T^I_\r(t)T^I(-t)$ belongs to $\A(I)$
if $t\geq 0$ as $\r(T^I(t)\cdot T^I(-t))$ is also localized in $I$.
Therefore $(X'\xi,\xi)=(X'\Q,\Q)$ if $X'\in\A(I')$.

Now $\xi$ is cyclic for $\A(I')$ if  $t\geq 0$, because $\xi =u\Q$
with $u=z(\r,t)$ a unitary in $\A(I)$.
By Proposition 2.10, the equation (2.10) gives
$$
\|e^{-\pi K_\r^I}T^I_\r(t)XT^I_\r(-t)\Q\|=
\|e^{-\pi K_\r^I}X\xi\|=
\|e^{-\pi K_\r^I}Xu\Q\|=
\psi_\r(Xuu^*X^*)=
\psi_\r(XX^*)
$$
showing (2.9) as desired.
\endproof
\proclaim\nofrills{\bf Proposition 2.12\usualspace} Let $\r=\oplus_k \r_k$ be a
direct sum of endomorphisms of $\A$ all localized in a given interval
$I$. Then
$$
(e^{-2\pi K^I_\r}\Q,\Q)= \sum_k (e^{-2\pi K^I_{\r_k}}\Q,\Q).
$$
\endproclaim
\proof
Let $v_k$ a family of isometries of $\A(I)$ such that
$\r=\sum_k v_k\r_k(\cdot)v_k^*$. Then
$$
e^{-2\pi K^I_\r}=\sum_k v_k e^{-2\pi K^I_{\r_k}}v_k^*
$$
hence by Proposition 2.10 we have
$$
(e^{-2\pi K^I_\r}\Q,\Q)=\sum_k (e^{-2\pi K^I_{\r_k}}v_k^*\Q,v_k^*\Q)
=\sum_k \psi_{\r_k}(v_k^*v_k)=\sum_k \psi_{\r_k}(1)=\sum_k d(\r_k)
$$
\endproof
\proclaim\nofrills{\bf Proposition 2.13\usualspace} Let $\r$ be as in
Proposition
2.10. Then $\r_I(\A(I))'\cap\A(I)$ is equal to the commutant
$\{\cup_{I_0}\r_{I_0}(\A(I_0))\}'$
of the representation $\r$.
\endproclaim
\proof
The proof is based on the arguments given in the proof of [\ref(GL2),
Theorem 2.3] that concerned the case
$\r$ had a priori finite index. In that context the proof relied
on the construction in [\ref(GL2), Corollary 2.5]  of a locally
normal faithful positive linear functional invariant  under
dilations and translations in the representation $\r$.  Proposion
2.11 provides us with such a functional $\tilde\psi_\r$ in our
setting, therefore, with obvious modifications, the rest of the proof
of [\ref(GL2), Theorem 2.3] is valid here.
\par\hfill\endproof
\proclaim\nofrills{\bf Lemma 2.14\usualspace} Let $\r$ be an endomorphism
of $\A$
localized in the interval $I$. If there exists a normal faithful
conditional $E$ expectation of $\A(I)$ onto $\r(\A(I))$, then $\r_I$
is a (possibly infinite) direct sum of irreducible endomorphisms of
$\A(I)$ with finite index. \endproclaim
\proof By conformal invariance we may assume that $I$ is the upper
semi-circle. We now use the formula
$$
\bar\r=j\cdot\r\cdot j
$$
for the conjugate sector $\bar\r$, where $j=\text{Ad}J$ with $J$ the
modular conjugation associated with the right semi-circle. Due to
its geometrical meaning, $j$ is an anti-automorphism of $\A(I)$,
so that $j\cdot E \cdot j$ is a normal faithful expectation onto
$\bar\r(\A(I))$. Now the inclusion $\bar\r(\A(I))\subset\A(I)$
is dual to the inclusion $\r(\A(I))\subset\A(I)$, hence the Lemma
follows by Proposition 2.6.
\endproof
\proclaim\nofrills{\bf Theorem 2.15\usualspace} Let $\r$ be an endomorphism
of $\A$
localized in the interval $I$.
Then  $(e^{-2\pi K^I_\r}\Q,\Q)<+\infty$ if and
only if $\r$ has finite index. Therefore the equality
$d(\r)=(e^{-2\pi K^I_\r}\Q,\Q)$ holds regardless
$d(\r)$ be finite or infinite.
\endproclaim
\proof We only have to show that if
$(e^{-2\pi K^I_\r}\Q,\Q)<+\infty$
then $\r$ has finite index. Now in this case Proposition 2.10
gives us a faithful positive normal linear functional on $\A(I)$
whose modular group leaves $\r(\A(I))$ globally invariant by
Proposition 1.1. By Takesaki theorem [\ref(Tak2)] we have a normal
faithful expectation onto $\r(\A(I))$, whence by Proposition 2.12
and Lemma 2.14 $\r$ is a direct sum of irreducible finite index
sectors $\r_k$. As for each $\r_k$ the formula $(e^{-2\pi
K^I_{\r_k}}\Q,\Q)=d(\r_k)$ holds true by Corollary 2.5, the results
follows by the additivity expressed in Proposition 2.12.
\endproof
Before concluding the section, we mention further applications of
the above methods to the analysis of superselection sectors with
infinite index [\ref(GL1), Section 11], in particular regarding the
positivity of the energy in these representations. This
matter will be discussed in [\ref(BCL)].
\medskip\np {\fsection 3. Hawking temperature in a charged
state and conditional  entropy.} \smallskip\np {\it 3.1 General
setting and a first expression.} Following the discussion made in the
introduction, we consider a Quantum Field Theory on the Minkowski
space $\Bbb R^4$, identify the Rindler space-time $W$ with a wedge
region in $\Bbb R^4$, and look at $W$ in analogy with the
Schwarzschild space-time.

 For convenience we fix the
Lorentz frame so that
$$
W=\{x\in \Bbb R^4,\, x_1>|x_0|\}
$$
and denote by $\L_W(t)$ the corresponding one-parameter group of
pure Lorentz transformations in the $x_1$-direction:
$$
\L_W(t) = \left( \matrix
\text {ch}(t) & \text {sh}(t) & 0 & 0\\
\text {sh}(t) & \text {ch}(t) & 0 & 0\\
0 & 0 & 1 & 0\\
0 & 0 & 0 & 1
\endmatrix \right ).
$$
Let $\A(\O)$ be the von Neumann algebra on the Hilbert space $\H$
of the observables localized within the region $\O$ of the Minkowski
space. Let $U$ denote the unitary covariant, positive energy,
representation of the Poincar\`e group $\Ppo$ on $\Cal H$ and $\Q$
the vacuum vector.

We assume the  local algebras to be generated by
a Wigthman field [\ref(SW)], in order to have the
Bisognano-Wichmann theorem that identifies the Tomita-Takesaki
modular operator $\D$ and the modular conjugation $J$ associated
with $(\A (W),\Q )$:
$$
\D_W^{it}= U(\L_W(-2\pi t)), \, t\in \Bbb R,
$$
and $J$ is the PCT anti-unitary composed with the unitary
implementation of the change of sign of the $x_2,x_3$-coordinates.
Therefore $U(\L_W(t))$ implements a one-parameter automorphism
group $\a_t$ of $\A(W)$ that satisfies the Kubo-Martin-Schwinger
equilibrium condition at inverse temperature $\b=2\pi$  with respect
to the restriction of the vacuum state $\o = ( .\Q,\Q)$ to
$\A(W)$\footnote{One may start with a  {\it modular covariance}
condition and encode the space-time symmetries intrinsically into
the net structure [\ref(BGL2)].}; in other words, by restriction
to $\A(W)$, the pure ground state  $\o$ becomes faithful (by the
Reeh-Schlieder theorem) and thermal for the geodesic evolution on
the Rindler space provided boost transformations.

As already explained, there is a relation of this
setting with the Hawking and the Unruh effects, first noted by Sewell
[\ref(Sew1)]. The space-time $W$ can be identified with the outside
region of a black hole. Then the observable algebra for the
background system of the black hole is $\A(W)$, the
corresponding proper Hamiltonian is
$$
H=aK=-i{\text{d}\over\text{d}t}U(\L_W(at))_{|t=0}
$$
where $a$ is the surface gravity of the black hole, and
the dynamics in the Heisenberg picture is given by
$$
\a_{at}(X)=e^{iHt}Xe^{-iHt},\quad X\in\A(W).
$$
Accordingly $\o_{|\A(W)}$ is a KMS state (i.e. Gibbs state at
infinite volume [\ref(HHW)]) at inverse  Hawking temperature
$\b={a\over 2\pi}$. We refer to [\ref(Haa),\ref(SV)] for more
details and further literature.

We shall consider the black hole as a heat reservoir for its
background system and treat the latter as an open
system.

Because of the particle production due to Hawking effect, the
background system undergoes a change in its quantum numbers,
namely the system goes in different superselection sectors, and we
shall consider the thermal equilibrium charged state corresponding
to a given sector.

We thus consider a superselection sector, namely the unitary
equivalence class of a representation $\r$ of the quasi-local
C$^*$-algebra $\A$, the norm closure  $\cup \A(\O)^-$ of the union
of all local algebras associated to bounded regions $\O$. The
representation $\r$ is assumed to be localizable in each space-like
cone $\S$\footnote{ This class exahusts all the
traslation covariant, positive energy representations with an
isolated mass shell [\ref(BF)], but possibly not
charges with long range interactions.}, namely  $\r$ and the identity
(vacuum) representation
have unitarily equivalent restrictions to the $C^*$-algebra
$\cup\{A(\O),\O\subset\S',\O\,\text{bounded}\}^-$ generated by the
local observables in $\S'$.

We may then
assume $\S\subset W$ and, by identifying the representation Hilbert
spaces, $\r$ to act as the identity on
 $\A(\O)$ if $\O\subset \S'$, namely $\r$ is a DHR localized
endomorphism [\ref(DHR)]. By wedge duality (consequence of the
Bisognano-Wichmann theorem)
 $$
 \A(W') = \A(W)',
 $$
 therefore $\r$ restricts to a normal endomorphism of $\A(W)$,
 still denoted by $\r$ (more precisely $\r$ restrict to $\A(W)\cap\A$
 and has a normal extension to $\A(W)$).

We assume that $\r$ is irreducible and Poincar\`e covariant with
positive energy-momentum, namely there exists
a unitary representation $U_\r$  of
universal covering group $\Sp$ of the  the Poincar\`e group such
that
$$ U_\r(g)\r(X)U_\r(g)^* = \r(U(g)XU(g)^*),\quad X\in\A,\,
g\in \Sp . \eqno(3.1)
$$
The covariance is automatic under general conditions [\ref(GL1)].

As shown in [\ref(DHR)], the notion statistics is intrinsically
associated with $\r$, in particular the statistical dimension
$d(\r)$ is defined and turns out to be a positive integer or
$+\infty$. We shall assume $d(\r)<\infty$. By the index-statistics
theorem [\ref(L1)]
$$
\text{Ind}(\r)=d(\r)^2,
$$
where $\text{Ind}(\r)$ is the minimal index of $\r_{|\A(W)}$,
so we may equivalently assume that the restriction of $\r$ to
$\A(W)$ has finite index.

The representation $U_\r$ giving the
covariance  is uniquely defined by formula (3.1): since $\r$ is
irreducible any other representation would differ from $U_\r$ by a
one-dimensional character of $\Spo$, that has to be trivial because
$\Spo$ has no non-trivial finite-dimensional unitary representation.

Now
$$
z(\r,g)=U_\r(g)U(g)^*
$$
is a Ad$U(g)$-cocycle and is localized, in particular $z(\r,g)$
belongs to $\A(W)$ if also $\r\cdot\text{Ad}U(g)$ is
localized in $W$. In particular $z(\r,\L_W(t))$ is a $\a_t$-cocycle
localized in $W$. As a consequence we have:
\proclaim\nofrills{\bf Lemma 3.1\usualspace}
$\a^\r_t(X)=U_\r(\L_W(t))XU_\r(\L_W(-t)),\, X\in\A(W)$ defines a
one-parameter  automorphism group $\a_t^\r$ of $\A(W)$. \endproclaim
\proof
We have
$$
\align
\a^\r_t(\A(W)) &= z(\r,\L_W(t))\a_t(\A(W))z(\r,\L_W(t))^* \\
&= z(\r,\L_W(t))\A(W)z(\r,\L_W(t))^*=\A(W)
\endalign
$$
because $z(\r,\L_W(t))$ belongs to $\A(W)$.
\endproof
The one-parameter automorphism group $\a^\r_{at}$ is the dynamics
of our system in the sector $\r$ and the corresponding proper
Hamiltonian is given by $$
H_\r = aK_\r = -i{\text{d}\over\text{d}t}U_\r(\L_W(at)){_|}_{t=0}
$$
Theorem 2.1, or equivalently formula (2.4), has
its version here, by an analogous proof,
$$
d(\r)=(e^{-\b H_\r}\Q,\Q)_{|\b ={2\pi\over a}} =
(e^{-2\pi K_\r}\Q,\Q)\eqno(3.2)
$$
where $\Q$ is the vacuum vector or any other cyclic vector for
$\A(W)$ such that $(\,\cdot\Q,\Q)$  coincides with the vacuum state
$\o$ on $\A(W')$.

As in Proposition 2.8 we have a normal faithful state of $\A(W)$
given by
$$
\f_\r(XX^*)= d(\r)^{-1}\|e^{-\pi K_\r}X\Q\|^2,\,
X\in\A(W).\eqno(3.3) $$
\proclaim\nofrills{\bf Lemma 3.2\usualspace} If $\r$ is irreducible
then the one-parameter automorphism group $\a_t^\r$ is ergodic on
$\A(W)$, namely its fixed points are the scalars. \endproclaim
\proof The proof is similar to the one given in [\ref(GL2)] in the
case of a conformal theory. Details will be given somewhere else.
\endproof
Next we show that the system in the sector $\r$ admits a thermal
equilibrium state at the same Hawking inverse temperature
$\b=2\pi/a$.
\proclaim\nofrills{\bf Theorem 3.3\usualspace}  $\a_{at}^\r$ admits a
unique normal KMS state $\f_\r$ at inverse temperature
 $\b={2\pi\over a}$.  The state $\f_\r$ is given by the equation
(3.3) or equivalently by
 $$
 \f_\r=\o\Phi_\r\eqno(3.4)
 $$
 where $\Phi_\r$ is the minimal left inverse of $\r$ on $\A(W)$
and, if $\r$
 is irreducible, $\f_\r$ is the unique normal $\a^\r$-invariant
state of $\A(W)$.
 If $\b\neq {2\pi\over a}$, no $\a_{at}^\r$-KMS normal state
exists. \endproclaim
\proof The situation is similar to the one discussed in
the previous section. Again, relying on on formula (3.2) and
Proposition 1.7, we see that the Connes cocycle
$(D\f_\r:D\o_{|\A(W)})_t$, where $\f_\r$ is defined by eq.
(3.4),  is
equal to $d(\r)^{-it}z(\r,\L_W(-2\pi t))$, thus  the modular group
of $\f_\r$ is given by
$$ \s_t^{\f_\r} =
\a^\r_{-2\pi t} $$
i.e. $\f_\r$ is $\a_{at}^\r$-KMS at $\b=2\pi/a$. The
non-existence of normal states at different temperatures is an
immediate
consequence of the outerness of $\s^{\f_\r}$  (cf.
[\ref(Con1)]), since  $\A(W)$ is a type $III_1$ factor (see
[\ref(L0)]).

The uniqueness of $\f_\r$ as a normal $\a^\r$-invariant state
is equivalent to the ergodicity of $\a^\r$, hence a consequence of
Lemma 3.2.
\endproof
Now a finite volume consideration (see Appendix B) suggests
that to regard $(e^{-\b H_\r}\Q,\Q)$ as the ratio of the (here
undefined) partition functions $Z_0(\b)$ of the state $\o$ and
$Z_\r(\b)$ of the state $\f_\r$, namely
$$
\log(e^{-\b H_\r}\Q,\Q)=\log Z_0(\b)-\log Z_\r(\b),\eqno(3.5)
$$
whence  we expect the quantity
$$
F(\o|\f_\r)=-\b^{-1}\log(e^{-\b H_\r}\Q,\Q)
$$
to represent the increment of the free energy between
$\o$ and $\f_\r$.
We shall see the above formula to hold true in a precise sense.

We define the {\it relative
free energy}   $F(\o|\f_\r)$ between the states
 $\o$ and $\f_\r$ by
$$
F(\o|\f_\r) = \f_\r(H_\r) - \b^{-1}S(\o|\f_\r) \eqno (3.6)
$$
where
$S(\o|\f_\r)=S(\o_{|\A(W)}|\f_\r)$ is the Araki relative entropy of
the two states on $\A(W)$,
$$
S(\o|\f_\r)
= -(\log\D_{\Q,\x_\r}\x_\r,\x_\r);\eqno(3.7)
$$
here  $\D_{\Q,\x_\r}$ is Araki's relative modular
operator  of $\A(W)$ associated with  the two cyclic
separating vectors $\Q$ and $\x_\r$, where $\x_\r$ is any
cyclic vector such that $\f_\r=(\cdot\ \x_\r,\x_\r)$ on $\A(W)$. In
particular we may assume  $\xi_\r$ to belong to the natural cone
associated with $(\A(W),\Q)$.

The quantity $\f_\r(H_\r)=(H_\r \xi_\r,\xi_\r)$ in $(3.6)$ represents the
{\it relative mean
energy} in between $\o$ and $\f_\r$, indeed according to formula
(3.5) this has to be given, anticipating Proposition
3.5, by
$$
\align
-{\text{d}\over\text{d}\b}\log(e^{-\b H_\r}\Q,\Q)&=
{(e^{-\b H_\r}H_\r\Q,\Q)\over(e^{-\b H_\r}\Q,\Q)}
={d(\r)(H_\r\D_{\Q,\xi_\r}^{1/2}\Q,\D_{\Q,\xi_\r}^{1/2}\Q)
\over(e^{-\b H_\r}\Q,\Q)}\\
&=(H_\r J\D_{\Q,\xi_\r}^{1/2}\Q,J\D_{\Q,\xi_\r}^{1/2}\Q)
=(H_\r \xi_\r,\xi_\r)=\f_\r(H_\r),
\endalign
$$
where $J$ is the  modular conjugation of both $\Q$ and $\xi_\r$ and
we have set $\b=2\pi/a$ and applied formula (3.2).

More directly one may define the relative mean energy by the formal
expression
$$
\tilde\f_\r(H_\r)=\tilde\f_\r(H_\r - H)
$$
where $H_\r - H$ is the relative Hamiltonian,
 $\tilde\f_\r=(\cdot\xi_\r,\xi_\r)$  and one sets
$\tilde\f_\r(H)=0$ motivated by the
fact that $(e^{-iHt}\xi_\r,\xi_\r)$ is a real even function
(because $\D_{\Q}^{it}$
preserves the natural cone) picking a maximum at $t=0$.

The relative entropy $S(\o|\f_\r)$ is always non-negative, but it
may be equal to $+\infty$, as no volume renormalization has been made
(cf. Appendix B); also the relative mean energy $\f_\r (H_\r)$ may be
infinite, but we shall show that the relative free
energy between $\f_\r$ and $\o$ is finite, so in particular
$\f_\r(H_\r)$ shall be bounded below.

Formula (3.6)
will have the obvious rigorous meaning as
$$
F(\o|\f_\r)=(H_\r+\b^{-1}\log\D_{\Q,\xi_\r}\xi_\r,\xi_\r).
$$
\proclaim\nofrills{\bf Theorem 3.4\usualspace} The relative free
energy between  the thermal equilibrium states $\o$ and $\f_\r$ is
proportional to the Connes-St\oe rmer entropy of the sector $\r$:
$$
F(\o|\f_\r)=-{1\over 2}\b^{-1}S_c(\r).
$$
\endproclaim
\np Here $S_c(\r)$ denotes the conditional entropy $H(\A(W)|\r(\A(W))$
(see (1.17), (1.18))
$$
S_c(\r)=\sup_{(\f_i)}\{\sum_{i=1}^n S(\f_\r|\f_i) -
S(\f_\r\cdot\r|\f_i\cdot\r)\}
$$
where ${(\f_i)}$ varies among the sets of finitely many normal
positive linear  functionals $\f_1, \f_2,\dots \f_n$ of $\A(W)$ such
that $\sum_{i=1}^n \f_i =\f_\r$.

We note the extensive property of $S_c$:
$S_c(\r_1\r_2)=S_c(\r_1)+S_c(\r_2)$ [\ref(L2)].
\proclaim\nofrills{\bf Proposition 3.5\usualspace} We have
$$
 \b H_\r=2\pi K_\r=-\log\D_{\Q,\x_\r} -  {1\over
2}S_c(\r).\eqno(3.8) $$\endproclaim
\proof
By Araki's formula
$$(D\f_\r:D\o_{|\A(W)})_t =
\D^{it}_{\Q,\x_\r}\D_\Q^{-it}
$$
therefore
$$u_W(\r,t) = (Dd(\r)\f_\r:D\o_{|\A(W)})_t
=d(\r)^{it}\D^{it}_{\Q,\x_\r}\D_\Q^{-it}.$$
On the other hand by our formula
$$u_W(\r,t) = z(\r,\L_W(-2\pi t)) = e^{i2\pi K_\r t}e^{-i2\pi Kt},$$
 therefore $$
e^{-i2\pi K_\r t}e^{i2\pi Kt} = d(\r)^{it}\D^{it}_{\Q,\x_\r}\D_\Q^{-it}
$$
and as $\D_\Q^{it} =U(\L_W(-2\pi t))=e^{-iKt}$ we see that
$$
e^{-i2\pi tK_\r } = d(\r)^{it}\D_{\Q,\x_\r}^{it}
$$
so the proposition is obtained by differentiating this expression
at $t=0$.

An alternative argument will appear in Lemma 3.10.
\endproof
\demo\nofrills{\it Proof of Theorem 3.4.\usualspace} By evaluating
on $\tilde\f_\r$ both sides of formula (3.7) we have
$$
 \b(H_\r\x_\r,\x_\r) =-(\log\D_{\Q,\x_\r}\x_\r,\x_\r) -  \log d(\r).
 $$
 On the other hand  $d(\r)$ is the square root of the minimal index
 of $\r_{|\A(W)}$ [\ref(L1)], thus by the Pimsner-Popa equality
 (1.18) it follows that $\log d(\r)={1\over 2}S_c(\r)$
hence proving the theorem.
\endproof
\proclaim\nofrills{\bf Corollary 3.6\usualspace} The possible
values of the relative free energy with initial state $\o$ are
$$
F(\o|\f_\r) = -{1\over 2}\b^{-1}\log(n), \quad n=1,2,3,\dots
$$
\endproclaim
\proof Immediate by the DHR theorem [\ref(DHR)] to the effect that
the statistical dimension is a positive integer or $+\infty$.
\endproof
\np Therefore the integer $n$, expressing the order of the
parastatistics in [\ref(DHR)], here appears  as a quantum number
labeling the relative free energy levels.

In low space-time dimensions the quantization of the conditional
entropy is less restrictive. By Jones theorem [\ref(Jon)] and the
results in [\ref(L2),\ref(Reh)] we have however restrictions for the
possible values of the relative free energy associated with a planar
black hole:
\proclaim\nofrills {\bf Corollary
3.7\usualspace} In low dimensions the possible values of
$e^{-\b F(\o|\f_\r)}$ are restricted to
$4\cos^2({\pi\over n})$ in the interval $(0,4)$. No value in
 $(4,5)$ is possible. In the interval $(5,6)$ only 3 values are
possible. \endproclaim
\proof The first assertion follows from [\ref(Jon)], because
$e^{-\b F(\o|\f_\r)}$ is an index. The rest of the statement is
a consequence of the further restrictions on the index values due
to the occurrence of the braid group symmetry [\ref(L2),\ref(Reh)].
\endproof \smallskip
\np{\it 3.2 The increment of the free energy between
arbitrary thermal equilibrium states.}
Let $\s$ be another
endomorphism localized in $W$ and $\xi_\s$ the cyclic separating
vector for $\A(W)$ such that
$$ (X\xi_\s,\xi_\s) = \f_\s(X), \quad X\in
\A(W) $$
lying in the natural cone associated with $(\A(W),\Q)$.

To extend the definition (3.6) for the free relative
energy to the case the initial state is an arbitrary thermal state
$\f_\s$, we note first that the formal relative Hamiltonian
between $\f_\s$ and $\f_\r$ is $H_\r-H_\s$ and hence the relative
mean internal energy should be formally given by
$$
\tilde\f_\r(H_\r-H_\s)=
\tilde\f_\r(H_\r-H_\s-H)=\tilde\f_\r(H_\r+H_{\bar\s}-H).
$$
Here the conjugate
charge  given by $\bar\s=j\cdot\s\cdot j$ (see [\ref(GL1)])
 is localized in $W'$ and
$H_{\bar\s}=JH_\s J=-H_\s$.

These premises and the following Lemma will motivate the definition
(3.9) below.
\proclaim\nofrills{\bf Lemma 3.8\usualspace} We have
$$e^{itH_{\r\bar\s}}=e^{itH_\r}e^{-itH}e^{itH_{\bar\s}}.$$
\endproclaim \proof
By the cocycle property
$$z(\r\bar\s,\L_W(t))=
\r(z(\bar\s,\L_W(t)))z(\r,\L_W(t))
=z(\bar\s,\L_W(t))z(\r,\L_W(t))
$$
because $z(\bar\s,\L_W(t))$ is localized in $W'$ and $\r$
acts identically on $\A(W')$ so the Lemma is obtained by multiplying
on the right by $e^{-itK}$ the above expression.
\endproof
We thus
define the relative free energy between $\f_\s$ and $\f_\r$ by
$$
F(\f_\s|\f_\r)=
\tilde\f_\r(H_{\r\bar\s})-\b^{-1}S(\f_\s|\f_\r)\eqno(3.9)
$$
and we give to this expression a rigorous meaning as done for the
expression (3.6).

Alternatively, extending the considerations in the previous
subsection, we could interpret directly
$\log(e^{-\b H_{\r\bar\s}}\xi_\s,\xi_\s)$
as the increment of the logarithm of the partition function between
 the states $\f_\s$ and $\f_\r$, leading to the expression
$$
F(\f_\s|\f_\r)= -\b^{-1}\log(e^{-\b H_{\r\bar\s}}\xi_\s,\xi_\s).
$$
\proclaim\nofrills{\bf Theorem 3.9\usualspace} The relative free energy is given
by
$$
F(\f_\s|\f_\r)={1\over 2}\b^{-1}(S_c(\s)-S_c(\r)).
$$
\endproclaim
\np As a consequence $e^{-\b F(\f_\s|\f_\r)}$ is equal to the
rational number $d(\r)\over d(\s)$.
\proclaim\nofrills{\bf Lemma 3.10\usualspace} Let $\r$
be as above, $\r'$ an endomorphism localized in $W'$ and
$\f_{\r'}=\o\cdot{\Phi_{\r'}}_{|\A(W')}$, where $\Phi_{\r'}$ is
minimal left inverse of $\r'$. Then
$$
\D(\f_{\r'}|\f_\r) ={d(\r)\over d(\r')}e^{-2\pi K_{\r\r'}}
$$
where $\D(\cdot|\cdot)$ denotes the Connes spatial derivative.
\endproclaim
\proof
Setting $\o'=\o_{|\A(W')}$ and using [\ref(Con0)] one has
$$\align
\D(\f_{\r'}|\f_\r)^{it}&=(D\f_{\r'}:D\o')_t\D(\o'|\f_\r)^{it}\\
&=d(\r)^{it}(D\f_{\r'}:D\o')_t e^{-i2\pi tK_\r}\\
&=d(\r)^{it}d(\r')^{-it}e^{-i2\pi tK_{\r'}}
e^{i2\pi tK}e^{-i2\pi tK_\r}\\
&=d(\r)^{it}d(\r')^{-it}z(\r',\L_W(-2\pi t))z(\r,\L_W(-2\pi
t))e^{-i2\pi tK}\\
&=d(\r)^{it}d(\r')^{-it}\r(z(\r',\L_W(-2\pi
t)))z(\r,\L_W(-2\pi t))e^{-i2\pi tK}\\
&=d(\r)^{it}d(\r')^{-it}z(\r\r',\L_W(-2\pi t))e^{-i2\pi tK}\\
&=d(\r)^{it}d(\r')^{-it}e^{-i2\pi tK_{\r\r'}}.\endalign $$
where
we have used Proposition 2.5 in our context. \endproof
\proclaim\nofrills{\bf Corollary 3.11\usualspace} We have
$$
\D_{\xi_\s,\xi_\r} =
{d(\r)\over d(\s)}e^{-2\pi K_{\r\bar\s}}.
$$
\endproclaim
\proof Immediate by the above discussion and the relation
$\D_{\xi_\s,\xi_\r}$ $= \D(\f_\s\cdot\text{Ad}J|\f_\r)$,
where $\f_\s\cdot\text{Ad}J$ is the vector state
$(\cdot\xi_\s,\xi_\s)$ on $\A(W')$.
\endproof
\demo\nofrills{\it Proof of Theorem 3.9.\usualspace} By the above
Corollary we have $$
\b H_{\r\bar\s}=\b aK_{\r\bar\s} = -\log\D_{\xi_\s,\xi_\r}+
{1\over 2}(S_c(\s)-S_c(\r))
$$
and this clearly implies the desired relation.\endproof
\medskip\np
{\fsection Appendix A. Tensor categories and cocycles.}\smallskip\np
The purpose of this Appendix is to
shed light on  part of the mathematical structure  underlying
our results. Indeed a good part of our  results
depends only on the tensor categorigal structure  provided by the
superselection sectors and are therefore visible without a more
detailed description of the theory.

Let $\T$ be a strict $C^*$-tensor category.   We assume $(\iota,\iota)=\Co$,
where $\iota$ is the identity object and $(\cdot,\cdot)$ denotes the
intertwiner space. We refer to [\ref(LR)]
for the basic notions used here.

A basic and originating example for $\T$, appearing in
[\ref(DHR),\ref(DR)],    is obtained by taking $\en$, $M$ a unital
$C^*$-algebra with trivial centre, to be the set of objects, and arrows
between objects $\r$ and $\r'$  given by $$ (\r,\r') = \{T\in M, \,
T\r(x) = \r'(x)T, \,\forall x\in M\} $$ while the tensor product is
given by  the composition of maps
$$
\r\otimes\r' = \r\cdot\r'
$$
$$
T\otimes S = \r_2 '(T)S = S\r_1 '(T),\quad T\in (\r_1,\r_2),\, S
\in (\r_1 ',\r_2 ').
$$
The reader unfamiliar with abstract tensor categories might at first focuses
on this particular case.

Given an object $\rho$ of $\Cal T$, an object
$\bar \rho$ of $\Cal T$ is said to be a conjugate of $\r$ if there
exist  $R_\r \in (\iota, \bar \rho \rho)$ and $\bar R_\r \in (\iota,
\rho \bar \rho)$ such that  $$
\bar R_\r^*\otimes 1_\rho \circ 1_\rho\otimes R_\r=1_\rho ;
\quad R_\r^*\otimes 1_{\bar \rho} \,\circ \, 1_{\bar\rho}\,\otimes
\bar R_\r=1_{\bar \rho}.\eqno(A.1)
$$
We shall assume that each object $\r$ has a conjugate $\bar\r$ (this
is automatic in $\en$ [\ref(L2)] if $M$ is an infinite factor and $\r$
has finite index) and
shall refer to (A.1) as to the conjugate equation for $\r$ and
$\bar\r$. The equation (A.1) has then a standard solution, namely
one can choose multiples of isometries $R_\r$ and $\bar R_\r$ in
(A.1) so that $\|R_{\r}\|=\|\bar R_\r\|=\sqrt{d(\r)}$ is minimal.
This formula defines $d(\r)$, the dimension of $\r$ [\ref(LR)].

Now recall that given an arrow $T\in (\r,\r')$, the conjugate
arrow $T^{\bullet}\in (\bar\r,\bar\r')$ is defined by
$$
T^\bullet=1_{\bar \rho'} \otimes \bar R_{\rho}^* \circ 1_{\bar
\rho'}  \otimes T^* \otimes
1_{\bar \rho} \circ R_{\rho'} \otimes 1_{\bar \rho}\in
(\bar \rho, \bar \rho'),
$$
where $\bar R_{\r}$ and $R_{{\r}'}$ are multiples of
isometries in the standard
solution for the conjugate equations defining the conjugates
$\bar\r$ and $\bar\r'$.  The mapping $T \mapsto T^\bullet$ is
antilinear and enjoys in particular the  following properties
\smallskip \item {$a$)} $1^\bullet_\rho = 1_{\bar \rho},$\par
\item {$b$)} $S^\bullet \circ T^\bullet = (S \circ T)^\bullet$
\item {$c$)} $T^{\bullet *} = T^{*\bullet}.$
\smallskip
\np We shall say that $\a$ is an (anti-)automorphism of $\T$ if $\a$
is an invertible functor of $\T$ with itself, (anti-)linear on the
arrows, commuting with the $^*$-operation and preserving tensor
products. The action of $\a$ on the object $\r$ and on the arrow $T$
will be denoted by $\r^\a$ and $T^\a$.

Given an automorphism $\a$ of $\T$ a {\it cocycle} $u$ with respect
to $\a$ is a map $$
\r\in \text{ Obj}(\T)\to u(\r) \text{\ unitaries\ in\ }(\r,\r^\a)
$$
such that
\item{$a$)}$u(\r\otimes\r^\prime)=u(\r)\otimes u(\r^\prime)$
\item{$b$)}If $T\in(\r,\r^\prime)$ then the following diagram
commutes $$
\CD
\r            @>T>>        \r^\prime  \\
@Vu(\r)VV				             @VVu(\r^\prime)V  \\
\r^\a         @>>T^\a>      \r^{\prime\a}
\endCD \eqno(A.2)
$$
\proclaim\nofrills{\bf Proposition A.1\usualspace} Let $u$ be a  cocycle with
respect to $\a$ as above. Then $u(\bar \r)=u(\r)^\bullet$.
\endproclaim
\proof
 The proof is obtained similarly as in Proposition 1.5.
 \endproof
\proclaim\nofrills{\bf Lemma A.2\usualspace} Let $u$ be a  cocycle with
respect to $\a$ as above and $j$ be an
anti-automorphism of $\T$. Then  $\r\to u(\r^j)^j$
is a cocycle with respect to $j\a j^{-1}$.
\endproclaim
\proof The statement is checked by a direct verification.
\endproof
We now give  two uniqueness results that are at the basis of the
identifications of the covariance cocycle and the Connes cocycle in
this paper.
\proclaim\nofrills{\bf Lemma A.3\usualspace} With the notations in
Lemma A.2, assume that $j\a j^{-1}
= \a^{-1}$ and  $\r^{j}$ is a conjugate of $\r$ for all objects
$\r$. Then the cocycle $u$ with respect to $\a$ is unique.
\endproclaim
\proof If $u'$ is a cocycle with respect to $\a$ and $\r$ is
irreducible, then $u'(\r)=\mu(\r)u(\r)$ for some phase $\mu(\r)$.
By Proposition A.1 $\mu(\bar\r)=\overline{\mu(\r)}$, while by Lemma
A.2 $\mu(\bar\r)=\mu(\r)$, hence $\mu=1$  on the irreducibles,
thus always because a cocycle is determined by its value on the
irreducible objects. \endproof
Let now
$G$ be a group and $\a$ an {\it action of $G$ on $\T$}, namely a
homomorphism $g\to\a_g$ of $G$ into the automorphism group of $\T$
(for simplicity we omit topological assumptions).

For any $\r\in\T$ and $g\in G$, let $u(\r,g)$ be a unitary in
$(\r_g,\r)$ (where $\r_g\equiv\r^{\a_g}$). We shall say that $u$ is
a {\it two-variable cocycle} if:
\item{$a$)} For any fixed $g\in G$,
$u(\cdot,g)^*$ is a cocycle with respect to the automorphism
$\a_g$ \item{$b$)} For any fixed $\r\in\T$, $u(\r,\cdot)$ is a
$\a$-cocycle, namely $u(\r,gh)=u(\r,g)u(\r,h)^{\a_g}$.
\proclaim\nofrills{\bf Proposition A.4\usualspace} Let $u$ be a two-variable
cocycle as above. If $G$ is perfect (i.e. $G$ has no non-trivial
one-dimensional unitary representation), then $u$ is unique.
\endproclaim \proof
As in the proof of Lemma A.3, if $\r$ is an irreducible object, a
second two-variable cocycle would give rise to a phase
$\mu(\r,g)$ that, for a fixed $\r$, would be a one-dimensional
character of $G$, and thus had to be trivial.\endproof
\medskip\np{\fsection Appendix B. The relative free energy at
finite volume.} \smallskip\np A finite volume
computation  with  canonical distribution may
clarify the notion of relative free energy $F$ in (3.6). Let
 the Hamiltonians of the evolutions $\a^{(0)}$ and $\a^{(1)}$ be
given by positive selfadjoint operators $H_0$ and $H_1$, so that
$\a^{(k)}$ is implemented by $e^{itH_k}$ .
The
Gibbs state $\o_\b^{(k)}$ for $\a^{(k)}$ is given by
$$
\o_\b^{(k)}= \text{Tr}(\r_k\cdot)
$$
with density matrix
$$
\r_k = Z_k(\b)^{-1}e^{-\b H_k}
$$
where $Z_k(\b)=\text {Tr}(e^{-\b H_k})$ is the  partition
function.

Then
$$
F_k = \o_k(H_k) - \b^{-1}S(\r_k)=-\b^{-1}\log Z_k(\b)
$$
is the Helmholtz free energy in the state $\o^{(k)}$,
where the entropy  in state
$\o^{(k)}$ is given by
$$
S(\r_k)=-\text{Tr}(\r_k\log\r_k)
$$
Then the relative entropy is given by (see [\ref(Th)])
$$\align
S(\o_\b^{(0)}|\o_\b^{(1)})=& -\text{Tr}(\r_0\log \r_0 -
\r_0\log \r_1) =- \o_\b^{(0)}(\log \r_0 - \log \r_1)\\
 &=\b\o_\b^{(0)}(H_{\text{rel}}) +
\log Z_0(\b) - \log Z_1(\b)
\endalign$$
where
$H_{\text{rel}} = H_0 - H_1$
is the relative
Hamiltonian.

The relative free energy is thus given by
$$\align
F(\o_\b^{(0)}|\o_\b^{(1)}) & =  \o_\b^{(0)}(H_{\text{rel}}) -
\b^{-1}S(\o_\b^{(0)}|\o_\b^{(1)}) \\& =
\b^{-1}\log Z_0(\b) - \b^{-1}\log Z_1(\b) = F_1 -F_0.
\endalign
$$
Had we considered a gran canonical distribution on the Fock space,
the Hamiltonian for $\a_t^{(k)}$ would have been implemented
 by $e^{it(H_k-\mu_k N_k)}$, with $\mu_k$  the chemical potential and
$N_k$ the number operator, and the above expression for $F_k$
would have had accordingly modified.

We note  explicitely that $$
e^{-\b F(\o^{(0)}_\b|\o^{(1)}_\b)} = {\text{Tr}(e^{-\b H_1})\over
\text{Tr}(e^{-\b H_0})}
$$
providing evidence to the analogy between formulae (0.1)
and (0.2). \medskip\np
{\fsection Final comments.}\smallskip\np
As mentioned, the Rindler space-time is a good
approximation of the Schwarzschild space-time only near the horizon.
However a version of our results within the context of the Kruskal
extension of the Schwarzschild space-time should be possible, as a
version of the Bisognano-Wichmann theorem and a model independent
derivation of the Hawking temperature has been given in this setting
[\ref(Sew2)].

Another point to comment on is related to the use of the Minkowski
vacuum associated with Poincar\`e symmetries. As is known there
exists no vacuum state for a quantum field theory on a general
curved space-time. In such a general theory the relative free energy
could however be defined by consistency with the fusion rules of the
supeselection structure and it seems  that our results may be
achieved in wider contexts.

Concerning the expression (3.6) for the relative free energy, it
would be physically meaningful to derive it by a finite
volume approximation, where its expression is given in Appendix B.
To this end one should use the split property and the Noether
currents, see [\ref(BDopL)], and this approach might also be useful
for the here above discussed extension to more general curved
spacetimes.
Moreover the development of such techniques could bring up
to a model independent derivation of the formula (0.1).
\medskip \np {\bf Acknowledgements.} It is a pleasure to thank
Claudio D'Antoni, Bernard Kay, John E. Roberts and Rainer Verch for
various conversations.

\bigskip
\centerline{\bf References}
\medskip
\references
\end